\documentclass[pra,twocolumn,showpacs,preprintnumbers,amsmath,amssymb,floatfix,longbibliography]{revtex4-1}

\usepackage{graphicx}
\usepackage{dcolumn}
\usepackage{bm}
\usepackage{color}
\usepackage{mathrsfs}

\newcommand{\bra}[1]{\langle #1|}
\newcommand{\ket}[1]{|#1\rangle}

\newcommand{\affA}{\vspace{1ex}$^1$Zentrum f\"ur Optische Quantentechnologien, Universit\"at Hamburg, Luruper Chaussee 149, 22761 Hamburg, Germany}
\newcommand{\affB}{$^2$Institut für Quantenphysik, Universit\"at Hamburg, Luruper Chaussee 149, 22761 Hamburg, Germany}
\newcommand{\affC}{$^3$The Hamburg Centre for Ultrafast Imaging, Luruper Chaussee 149, 22761 Hamburg, Germany}

\usepackage{ulem}
\usepackage{soul}
\usepackage{physics}
\usepackage{multirow}

\newcommand{\addtwo}[1]{#1}

\newcommand{\addone}[1]{#1}

\usepackage{makecell}

\usepackage{hyperref}
\usepackage{xcolor}
\hypersetup{
    colorlinks,
    linkcolor={red!70!black},
    citecolor={green!65!black},
    urlcolor={blue!80!black}
}

\begin{document}

\title{Cooling dynamics of a free ion in a Bose-Einstein condensate}

\author{Lorenzo Oghittu$^1$, Juliette Simonet$^{1,2,3}$, Philipp Wessels-Staarmann$^{1,2,3}$, \\Markus Drescher$^{1,3}$, Klaus Sengstock$^{1,2,3}$, Ludwig Mathey$^{1,2,3}$, Antonio Negretti$^1$}
\affiliation{\affA, \affB, \affC}
\date{\today}

\begin{abstract}
    We investigate the dynamics of an ion moving through a homogeneous Bose-Einstein condensate (BEC) after an initial momentum is imparted. For this, we derive a master equation in the weak-coupling limit and Lamb-Dicke approximation for the reduced density matrix of the ion. We study the time evolution of the ion's kinetic energy and observe that its expectation value, identified as the ion temperature $T_\mathrm{ion}$, is reduced by several orders of magnitude in a time on the order of microseconds for a condensate density in the experimentally relevant range between $10^{13}\,\mathrm{cm}^{-3}$ and $10^{14}\,\mathrm{cm}^{-3}$. We characterize this behavior by defining the duration at half maximum as the time required by $T_\mathrm{ion}$ to reach half of its initial value, and study its dependence on the system parameters. Similarly, we find that the expectation value of the ion's momentum operator is reduced by nine orders of magnitude on the same timescale, making the ion's position converge to a final value. Based on these results, we conclude that the interaction with the bosonic bath allows for cooling and pinning of the ion by decreasing the expectation value of its kinetic energy and velocity, which constitutes a result of direct relevance for current atom-ion experiments.
\end{abstract}

\maketitle

\section{Introduction}
\begin{figure}
    \centering
    \includegraphics[width=.45\textwidth]{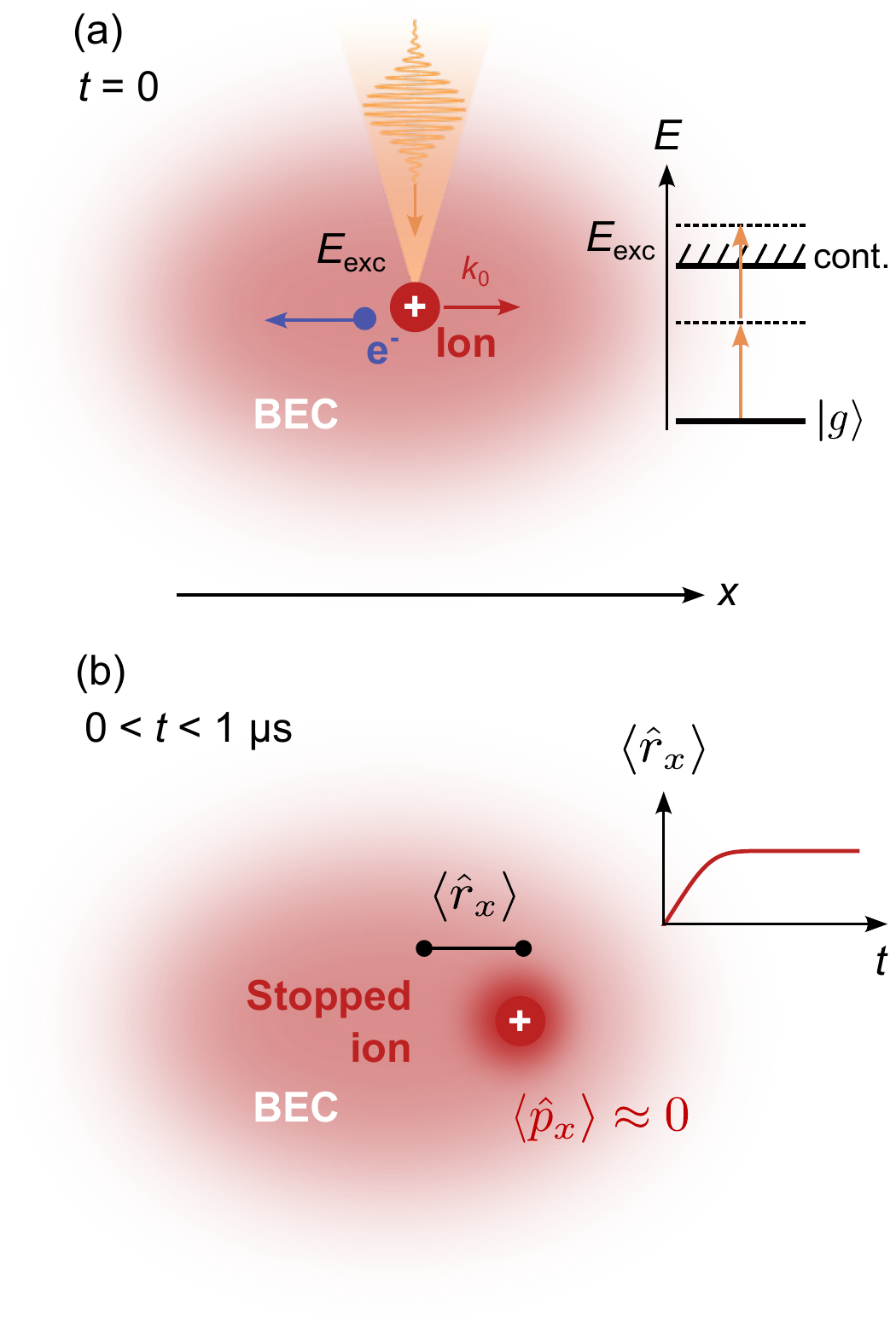}
    \caption{Proposal for cooling an ion in an ultracold bosonic gas. \textbf{(a)} An atom in a BEC is ionized by an ultrashort laser pulse via a non-resonant two-photon process. The excess energy $E_\mathrm{exc}$ of the ionization process determines the initial momentum $k_0$ of the ion in $x$ direction. \textbf{(b)} Subsequently, the ion is slowed down due to the atom-ion interaction arising from the polarizability of the atomic cloud. By deriving a master equation, we can extract the time-evolution of the expectation value of the ion’s position $\langle\hat{r}_x\rangle$ and show that it is pinned within a microsecond in a $^{87}$Rb BEC.}
    \label{fig:scheme}
\end{figure}

Quantum mixtures of ultracold atoms and ions have attracted the interest of an increasing part of the ultracold quantum matter community in the last few years. Combining the high controllability of trapped ions with the long coherence times of ultracold atomic systems, they provide a fertile platform for the study of both few- and many-body physics and their application to the advancement of quantum technologies arising from the long-ranged character of atom-ion interactions. Some of the most recent theoretical investigations include ab-initio quantum Monte Carlo and multi-configuration time-dependent Hartree methods for bosons as well as diagrammatic techniques for the analysis and characterization of polaronic states~\cite{Gregory2021,Bruun_PRL21,SchurerPRL17}. More recently, studies have also focused on how the interaction between two ions is mediated by the surrounding gas~\cite{Bruun_PRL22,Astrakharchik2023}, while proposals to exploit ions in ultracold gases as quantum simulators~\cite{BissbortPRL13,NegrettiPRB14, Michelsen2019,JachymskiPRR2020} or sensors~\cite{Oghittu_PRR22} have been put forward. We refer to Refs.~\cite{CoteAAMOP16,Tomza_RMP19} for an overview in the field. As far as experiments are concerned, most of the recent achievements involve the presence of external potentials that tightly trap the ion~\cite{Haerter2014}. In particular, sympathetic cooling was observed in such setups with the ion confined in radio-frequency traps~\cite{Gerritsma_NAT20,HirzlerPRA2020} or in optical dipole traps~\cite{SchmidtPRL2020,Schaetz_NAT21}. Similar systems were also employed in the observation and study of few-body processes and chemical reactions between ions and atoms~\cite{Denschlag_PRL12,RatschbacherNatPhys12,SikorskyNatCom18,Rios_MP21,Ozeri_NatPhys23}. On the other hand, experiments based on the ionization of Rydberg atoms~\cite{Kleinbach2018,Engel2018} have explored the scenario where no trap is present and the ion is driven by an external electric field, focusing on the transport properties of electrical charges inside a Bose-Einstein condensate~\cite{Meinert_PRL21,DieterlePRA2020} and the formation of molecules in Rydberg-atom-ion systems~\cite{Pfau_NAT22, Bosworth2023}. However, while the formation and behavior of neutral polarons both in the case of Fermi~\cite{SchirotzekPRL09,KoschorreckNature12,Kohstall_Nat12,ScazzaPRL17} and Bose environments~\cite{Hu2016,Jorgensen2016,Yan_Sci20} has made tremendous progress, the physics of mobile charged impurities in ultracold gases is at an earlier stage compared to its neutral analogous. This is due to the experimental challenges in reaching the ultracold regime involving only a few partial waves, due to the notorious micromotion~\cite{CetinaPRL12}. Theoretical challenges arise from the fact that the properties of the systems depend not only on the scattering length and effective range of the atom-ion potential, but also on the presence of the long-range tail of the interaction, preventing the use of the pseudopotential approximation~\cite{IdziaszekPRA07,GaoPRL10}.\\
Here, we study the quantum dynamics of a free, i.e., not trapped, ion moving inside a bosonic quantum gas with a finite initial momentum. Let us note that one-dimensional in-depth investigations of the quantum dynamics of the motional degrees of freedom of an ion both at zero and finite temperature interacting with matter waves confined in a double well have been carried out in Refs.~\cite{Joger2014,Ebgha_PRA19}. The ion-induced correlated dynamics of a bosonic system after ionization has been analyzed in Ref.~\cite{SchurerNJP15}, where the ion, however, has been treated as a static impurity.
Specifically, we resort to the master equation approach developed in Refs.~\cite{Idziaszek_PRA15, Oghittu_PRA21} to characterize the evolution of the expectation value of the ion's kinetic energy, velocity and position. Our study is motivated by the recent experimental advances involving untrapped ions in condensates~\cite{Meinert_PRL21,Simonet_NC21}, where we note that optical control of the ion movement in the atomic gas can be accomplished by means of optical traps as well~\cite{SchneiderNP10,Lambrecht2016}. In this work, we are inspired by the specific scenario that originates from the experiment reported by T. Kroker~\textit{et al.} in Ref.~\cite{Simonet_NC21}. As depicted in Fig.~\ref{fig:scheme}(a), a laser pulse ionizes some of the $^{87}$Rb atoms in a BEC within $215\,\mathrm{fs}$, hence instantly creating ions inside the bosonic gas with a finite initial kinetic energy determined by the excess energy of the ionization process. For this reason, we focus mostly on the case of the homonuclear system $^{87}$Rb$^+$/$^{87}$Rb, as this is the atomic species utilized in those experiments, but we also provide a brief analysis of the case of ions with a larger mass. We note that although in Refs.~\cite{Wessels_CP18,Simonet_NC21} the initial kinetic energy of the ion is on the order of a few microelectronvolts, this can be experimentally reduced by an order of magnitude. Fig.~\ref{fig:scheme}(b) illustrates that for the corresponding initial momentum the ion can be cooled and pinned within the BEC due to the long range atom-ion interaction arising from the polarizability of the atomic cloud. We characterize the cooling of the ion and find it to be remarkably robust against the initial ion velocity, density and temperature of the BEC as well as the mass ratio between the ion and the atoms.\\ 
\\
The paper is organized as follows: in Sec.~\ref{sec:system} we introduce the atom-ion interaction potential and the Hamiltonian describing the hybrid atom-ion system, while in Sec.~\ref{sec:ME} we derive the ionic master equation from which the equation of motion of the most relevant observables are obtained. In Sec.~\ref{sec:results} we analyse and discuss the results of the numerical simulations, while in Sec.~\ref{sec:exp} we discuss the experimental implications of the study. Finally, in Sec.~\ref{sec:summary} we summarize our findings and provide an outlook for future analysis.

\section{System}
\label{sec:system}
In this section, we briefly characterize the theoretical treatment of our system. For a more thorough description, we refer the interested reader to Refs.~\cite{Idziaszek_PRA15,Oghittu_PRA21}.

\subsection{Atom-ion interaction potential}
\label{sec:atom-ion_potential}
\begin{figure}
    \centering
    \includegraphics[width=.45\textwidth]{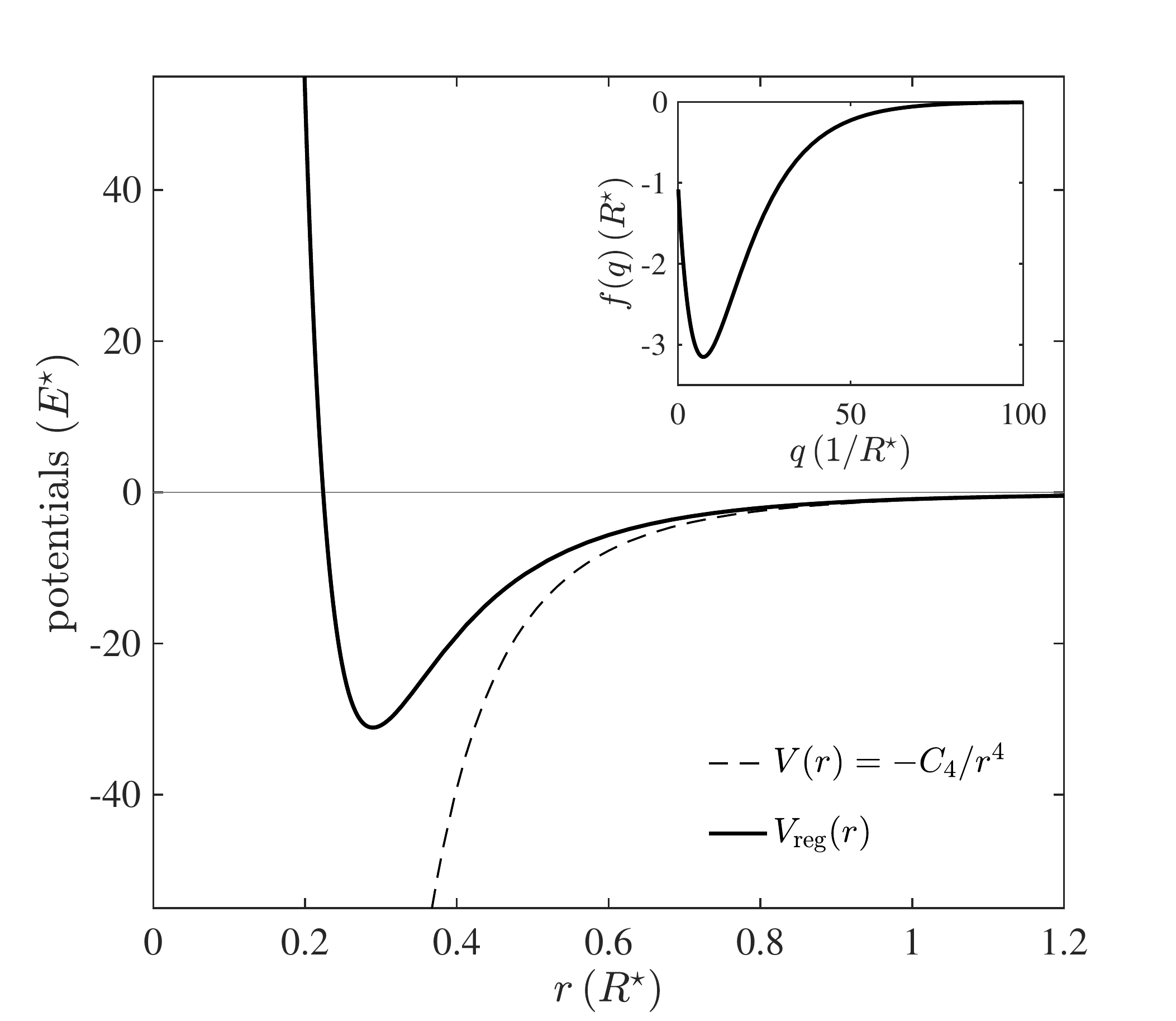}
    \caption{Main plot: Atom-ion interaction potential in units of $E^\star$ as a function of the atom-ion separation $r$ in units of $R^\star$. Dashed line: polarization potential. Solid line: regularized potential in Eq.~\eqref{eq:Vreg} with parameters $b=0.07797\,R^\star$, $c=0.2239\,R^\star$, \addtwo{corresponding to an atom-ion scattering length $a_{\mathrm{ai}}\simeq R^\star$ and a single two-body bound state with binding energy $E_\mathrm{BS}\simeq-1.43\,E^\star$}. Inset: scattering amplitude $f(q)$ [see Eq.~\eqref{eq:scattering_amplitude}] corresponding to the regularized potential in the main plot. Note that $f$ is in units of $R^\star$, whereas $q$ in units of $1/R^\star$.}
    \label{fig:potential}
\end{figure}
The interaction between a charged and a neutral particle depends on their separation $r=|\mathbf{r}|$. It is described asymptotically by the polarization potential $V(r)=-C_4/r^{4}$, where $C_4=\alpha e^2/(8\pi\epsilon_0)$ with $\alpha$ the static polarizability of the atom, $e$ the electron charge and $\epsilon_0$ the vacuum permittivity. This potential has the characteristic length $R^\star=\sqrt{2\mu C_4/\hbar^2}$ and energy $E^\star=\hbar^2/[2\mu(R^\star)^2]$, with $\mu$ as reduced mass. The value of $R^\star$ is much larger than the length scale of the van der Waals interaction between neutral particles and, for typical atom-ion systems, it is of the order of hundreds of nanometers. In particular, for the $^{87}$Rb/$^{87}$Rb$^+$ system we have $R^\star\simeq265.81\,\mathrm{nm}$ and $E^\star\simeq k_\mathrm{B}\cdot79\,\mathrm{nK}$ ($k_\mathrm{B}$ is the Boltzmann constant).\\
Due to the singularity of the polarization potential and the fact that we shall have to calculate its Fourier transform, we consider the following regularization \cite{Idziaszek_PRA15}
\begin{equation}
    V_\mathrm{reg}(r)=-C_4\frac{r^2-c^2}{r^2+c^2}\frac{1}{(r^2+b^2)^2},
    \label{eq:Vreg}
\end{equation}
\addtwo{where the energy spectrum and the atom-ion scattering length $a_\mathrm{ai}$ are controlled by the parameters $b$ and $c$ \footnote{\addtwo{Note that, to date, the short-range part of the atom-ion potential has not yet been experimentally characterized.}}}. The choice of the values of those parameters is discussed extensively in Ref.~\cite{Oghittu_PRA21}. An example of the potential is displayed in the main plot Fig.~\ref{fig:potential}.\\
The scattering amplitude in the first-order Born approximation is proportional to the Fourier transform of the potential, and is given by
\begin{equation}
    \begin{split}
        f(q)&=-\frac{\mu}{2\pi\hbar^2}\int_{\mathbb{R}^3}d\mathbf{r}\,e^{i\mathbf{q}\cdot\mathbf{r}}V_\mathrm{reg}(r)\\[1ex]
        &=\frac{c^2\pi(R^\star)^2}{(b^2-c^2)^2q}\bigg\{e^{-bq}\bigg[1+\frac{(b^4-c^4)q}{4bc^2}\bigg]-e^{-cq}\bigg\}.
    \end{split}
    \label{eq:scattering_amplitude}
\end{equation}
An example is shown in the inset of Fig.~\ref{fig:potential}, where $f(q)$ approaches zero for large momenta, while at $q R^\star \simeq7.37$ it exhibits a minimum. The expression of the scattering amplitude is used in the derivation of the master equation, as we shall discuss it in Sec.~\ref{sec:ME}.

\subsection{Hamiltonian}
\label{sec:Hamiltonian}
We consider a non-trapped ion of mass $M$ coupled to an ultracold bosonic gas with mass $m$, henceforth referred to as bath. The Hamiltonian is the sum of three terms: $\hat{H}=\hat{H}_\mathrm{ion}+\hat{H}_\mathrm{bath}+\hat{H}_\mathrm{int}$, with $\hat{H}_\mathrm{ion}=\hat{p}^2/(2M)$,
\begin{equation}
    \hat{H}_\mathrm{bath}=\int_{\mathbb{R}^3}d\mathbf{r}_b\,\hat{\Psi}_b^\dag(\mathbf{r}_b)\bigg[\frac{\hat{p}_b^2}{2m}+\frac{g}{2}\hat{\Psi}_b^\dag(\mathbf{r}_b)\hat{\Psi}_b(\mathbf{r}_b)\bigg]\hat{\Psi}_b(\mathbf{r}_b),
    \label{eq:H_bath}
\end{equation}
and
\begin{equation}
    \hat{H}_\mathrm{int}=\int_{\mathbb{R}^3}d\mathbf{r}_b\,\hat{\Psi}_b^\dag(\mathbf{r}_b)V_\mathrm{ib}(\mathbf{r}_b-\hat{\mathbf{r}})\hat{\Psi}_b(\mathbf{r}_b),
    \label{eq:H_int}
\end{equation}
where the subscript $b$ indicates the bosons of the bath, $\hat{\mathbf{r}}$ is the position operator of the ion, and $V_\mathrm{ib}$ represents the two-body potential between the ion and the particles of the bath. Moreover, we assume the bath to be confined in a box of length $L$ and its atoms to interact via contact potential with coupling strength $g=4\pi\hbar^2a_\mathrm{bb}^s/m$, $a_\mathrm{bb}^s$ being the \addtwo{three-dimensional (3D)} atom-atom $s$-wave scattering length.\\
The bosonic field operator can be written as an expansion around the condensate density $n_0=N_0/L^3$ ($N_0$ being the number of condensed particles) as
\begin{equation}
    \hat{\Psi}_b(\mathbf{r}_b)=\sqrt{n_0}+\delta\hat{\Psi}_b(\mathbf{r}_b),
    \label{eq:psi_b_expansion}
\end{equation}
where the fluctuations are described within Bogoliubov theory\addtwo{, i.e., }
\begin{equation}
\delta\hat{\Psi}_b(\mathbf{r}_b)=L^{-3/2}\sum_\mathbf{q}\Big(u_\mathbf{q}\hat{b}_\mathbf{q}e^{i\mathbf{q}\cdot\mathbf{r}_b}+v_\mathbf{q}\hat{b}_\mathbf{q}^\dag e^{-i\mathbf{q}\cdot\mathbf{r}_b}\Big),
    \label{eq:field_fluctuation}
\end{equation}
with $[\hat{b}_\mathbf{q},\hat{b}_{\mathbf{q}'}^\dag]=\delta_{\mathbf{q},\mathbf{q}'}$. \addtwo{By using Eq.~\eqref{eq:field_fluctuation}} we can rewrite the bath Hamiltonian as follows
\begin{equation}
    \hat{H}_\mathrm{bath}\approx E_0+\sum_\mathbf{q}\hbar\omega_\mathbf{q}\hat{b}_\mathbf{q}^\dag\hat{b}_\mathbf{q}
    \label{eq:H_b_Bogoliubov}
\end{equation}
with $E_0=gN_0^2/(2L^3)$ the condensate ground-state energy, and the phononic dispersion relation given by
\begin{equation}
    \addtwo{\epsilon(\mathbf{q})\equiv\,}\hbar\omega_\mathbf{q}=\sqrt{\bigg(\frac{\hbar^2q^2}{2m}\bigg)^2+\big(\hbar c_s q\big)^2},
    \label{eq:dispersion}
\end{equation}
where $c_s=\sqrt{gn_0/m}$ is the speed of sound of the gas. The amplitudes of the Bogoliubov modes are given by~\cite{LifshitzVol10}
\begin{equation}
    \begin{split}
        u_\mathbf{q}=&\sqrt{\frac{\hbar^2q^2/(2m)+gn_0}{2\hbar\omega_\mathbf{q}}+\frac{1}{2}}\\
        v_\mathbf{q}=&-\sqrt{\frac{\hbar^2q^2/(2m)+gn_0}{2\hbar\omega_\mathbf{q}}-\frac{1}{2}}.
    \end{split}
    \label{eq:Bogoliubov_amplitudes}
\end{equation}
Hence, the atomic density operator reads
\begin{equation}
    \hat{\Psi}_b^\dag(\mathbf{r}_b)\hat{\Psi}_b(\mathbf{r}_b)=n_0+\Delta\hat{n}(\mathbf{r}_b)
\end{equation}
and we can use the definition in Eq.~\eqref{eq:psi_b_expansion} to write the last term on the right hand side as
\begin{equation}
    \Delta\hat{n}(\mathbf{r}_b)=\delta\hat{n}(\mathbf{r}_b)+\delta^2\hat{n}(\mathbf{r}_b)
\end{equation}
with $\delta\hat{n}(\mathbf{r}_b)=\sqrt{n_0}[\delta\hat{\Psi}_b(\mathbf{r}_b)+\delta\hat{\Psi}_b^\dag(\mathbf{r}_b)]$ and $\delta^2\hat{n}(\mathbf{r}_b)=\delta\hat{\Psi}_b^\dag(\mathbf{r}_b)\delta\hat{\Psi}_b(\mathbf{r}_b)$. In our description \addtwo{we only consider the first of the two terms, thereby taking into account only the density fluctuations proportional to the square root of the condensate density $n_0$.} Let us note that the second order is related to the non-condensed part of the gas. As we pointed out in Ref.~\cite{Oghittu_PRA21}, its contribution becomes relevant when the gas temperature approaches the critical temperature of condensation from below, and is the only one contributing in the absence of condensation. Here, however, our analysis focuses on gas temperatures much lower than the critical temperature, allowing the contribution of the quadratic terms to be safely neglected. According to Eq.~\eqref{eq:field_fluctuation}, we have
\begin{equation}
    \begin{split}
        \delta\hat{n}(\mathbf{r}_b)=\sqrt{\frac{n_0}{L^3}}\sum_\mathbf{q}\bigg[&\big(u_\mathbf{q}+v^*_\mathbf{q}\big)\hat{b}_\mathbf{q}e^{i\mathbf{q}\cdot\mathbf{r}_b}\\
        &+\big(u_\mathbf{q}^*+v_\mathbf{q}\big)\hat{b}_\mathbf{q}^\dag e^{-i\mathbf{q}\cdot\mathbf{r}_b}\bigg].
    \end{split}
\end{equation}
At this stage let us remark that we assume that the condensate density is not affected by the presence of the ion and remains homogeneous. As recently shown in Ref.~\cite{Gregory2021}, however, the formation of many-body bound-states can change the bath density around the ion substantially. Such many-body bound-states are not included in the present study, since their formation remains negligible as long as no stimulated resonance processes occur~\cite{CotePRL02}. Under these assumptions, our open system approach is justified.\\
Finally, the interaction Hamiltonian becomes
\begin{equation}
    \begin{split}
        \hat{H}_\mathrm{int}&=\int_{\mathbb{R}^3}d\mathbf{r}_b\,V_\mathrm{ib}(\mathbf{r}_b-\hat{\mathbf{r}})\delta\hat{n}(\mathbf{r}_b)\\
        &=\hbar\sum_\mathbf{q}\Big(\hat{S}_\mathbf{q}\hat{b}_\mathbf{q}+\hat{S}^\dag_\mathbf{q}\hat{b}^\dag_\mathbf{q}\Big)
    \end{split}
    \label{eq:Froehlich}
\end{equation}
with
\begin{equation}
    \hat{S}_\mathbf{q}=\frac{\sqrt{n_0L^3}}{\hbar}\big(u_\mathbf{q}+v_\mathbf{q}^*\big)e^{i\mathbf{q}\cdot\hat{\mathbf{r}}}c_\mathbf{q}
\end{equation}
and
\begin{equation}
    c_\mathbf{q}=\frac{1}{L^3}\int_{\mathbb{R}^3}d\mathbf{y}\,e^{i\mathbf{q}\cdot\mathbf{y}}V_\mathrm{ib}(\mathbf{y}).
\end{equation}
Note that the coefficient $c_\mathbf{q}$ is related to the scattering amplitude $f(q)$ by 
\begin{equation}
    c_\mathbf{q}=-\frac{2\pi\hbar^2}{\mu L^3}f(q)
\end{equation}
As discussed in Sec.~\ref{sec:atom-ion_potential}, we model the two-body atom-ion potential $V_\mathrm{ib}$ with the regularization \addtwo{of} Eq.~\eqref{eq:Vreg}, whose scattering amplitude is given in Eq.~\eqref{eq:scattering_amplitude}.\\

\section{Impurity master equation}
\label{sec:ME}
In this section we derive a master equation for the reduced density matrix of the ion. We start from the master equation in the Born and Markov approximation for an impurity in a bosonic bath:
\begin{equation}
    \begin{split}
         \frac{d}{dt}\hat{\rho}(t)=-\frac{i}{\hbar}\comm{\hat{H}_\mathrm{ion}}{\hat{\rho}(t)}-\sum_\mathbf{q}\int_0^{t}d\tau\,\Omega_\mathbf{q}^2\,&\bigg\{\\
         +\big(n_\mathbf{q}+1\big)\comm{e^{i\mathbf{q}\cdot\hat{\mathbf{r}}}}{e^{-i\mathbf{q}\cdot\hat{\mathbf{r}}(t,\tau)}\hat{\rho}(t)}&e^{-i\omega_\mathbf{q}\tau}\\
         +\,n_\mathbf{q}\comm{\hat{\rho}(t)e^{-i\mathbf{q}\cdot\hat{\mathbf{r}}(t,\tau)}}{e^{i\mathbf{q}\cdot\hat{\mathbf{r}}}}&e^{-i\omega_\mathbf{q}\tau}\\
         +\big(n_\mathbf{q}+1\big)\comm{\hat{\rho}(t)e^{i\mathbf{q}\cdot\hat{\mathbf{r}}(t,\tau)}}{e^{-i\mathbf{q}\cdot\hat{\mathbf{r}}}}&e^{i\omega_\mathbf{q}\tau}\\
         +\,n_\mathbf{q}\comm{e^{-i\mathbf{q}\cdot\hat{\mathbf{r}}}}{e^{i\mathbf{q}\cdot\hat{\mathbf{r}}(t,\tau)}\hat{\rho}(t)}&e^{i\omega_\mathbf{q}\tau}\bigg\}.
    \end{split}
    \label{eq:BEC_ME}
\end{equation}
\addtwo{Here, we defined}
\begin{equation}
    \Omega_\mathbf{q}^2=\frac{|u_\mathbf{q}+v_\mathbf{q}|^2}{\hbar^2}|c_\mathbf{q}|^2n_0L^3,
\end{equation}
\addtwo{while $n_\mathbf{q}=[e^{\beta(\epsilon(\mathbf{q})-\mu_\mathrm{B})}-1]^{-1}$ is the Bose-Einstein occupation number based on the averages over the thermal state of the bath $\hat{B}_0$ [see also Eq.~(\ref{eq:B0})]}
\begin{equation}
    \mathrm{Tr}_\mathrm{b}\big\{\hat{b}_\mathbf{q}^\dag\hat{b}_{\mathbf{q}'}\hat{B}_0\big\}=n_\mathbf{q}\,\delta_{\mathbf{q},\mathbf{q}'},
\end{equation}
\addtwo{with $\mu_\mathrm{B}$ the chemical potential of the bosonic gas at temperature $T_\mathrm{gas}$, and $\beta=1/(k_\mathrm{B}T_\mathrm{gas})$.} We note that Eq.~\eqref{eq:BEC_ME} corresponds to the first line of Eq.~(41) in Ref.~\cite{Oghittu_PRA21} and can be applied to any kind of impurity in interaction with a bath of bosonic atoms by specifying the scattering amplitude in the definition of $c_\mathbf{q}$ and the equation of motion of the impurity $\hat{\mathbf{r}}(t,\tau)$. For a detailed derivation we refer to Ref.~\cite{Oghittu_PRA21}. 

\subsection{Lamb-Dicke approximation}
\label{sec:LDapprox}
In order to render the master equation numerically treatable, we perform the Lamb-Dicke approximation to further simplify it. Such an approximation is based on the assumption that the average wavelength of the atoms in the bosonic bath, corresponding to the de Broglie wavelength $\lambda_\mathrm{dB}(T_\mathrm{gas})$, is much larger than the \addtwo{spatial extension} of the ion, namely the width of the associated wave packet. \addtwo{The validity of this requirement is discussed in Appendix~\ref{app:LD}, while here we proceed with the derivation of the master equation.} In the Lamb-Dicke regime, we Taylor expand the products of exponential functions containing $\mathbf{q}\cdot\hat{\mathbf{r}}$ and $\mathbf{q}\cdot\hat{\mathbf{r}}(t,\tau)$ and keep the terms up to second order. For instance, the first commutator in Eq.~\eqref{eq:BEC_ME} can be written as
\begin{equation}
    \begin{split}
        &\comm{e^{i\mathbf{q}\cdot\hat{\mathbf{r}}}}{e^{-i\mathbf{q}\cdot\hat{\mathbf{r}}(t,\tau)}\hat{\rho}(t)}\simeq\\
        &i\comm{\mathbf{q}\cdot\hat{\mathbf{r}}}{\hat{\rho}(t)}+\comm{\mathbf{q}\cdot\hat{\mathbf{r}}}{\mathbf{q}\cdot\hat{\mathbf{r}}(t,\tau)\hat{\rho}(t)}-\frac{1}{2}\comm{(\mathbf{q}\cdot\hat{\mathbf{r}})^2}{\hat{\rho}(t)}.
    \end{split}
    \label{eq:LD_expansion}
\end{equation}
However, due to the assumed spherical symmetry of the bath, the first term on the right hand side of Eq.~\eqref{eq:LD_expansion} is zero after the sum over $\mathbf{q}$ is taken, and so are the terms containing odd powers of $q_x$, $q_y$ or $q_z$. Hence, the directions are decoupled and the contribution from the first commutator reads $\sum_\xi(\comm{\hat{r}_\xi}{\hat{r}_\xi(t,\tau)\hat{\rho}(t)}-[\hat{r}_\xi^2,\hat{\rho}(t)]/2)$, $\xi=x,y,z$.\\
We now explicitly substitute the equation of motion of the free ion $\hat{r}_\xi(t,\tau)=\hat{r}_\xi-(\hat{p}_\xi/M)\tau$ and perform the time integration. We note that the latter is performed analytically in the present study, which is in contrast with the usual approach in the literature~\cite{CarmichaelQOBook} and with the previous works~\cite{Idziaszek_PRA15,Oghittu_PRA21}, where the limit $t\to\infty$ has been taken. For further details we refer to the Appendix~\ref{app:ME}. Similarly to Ref.~\cite{Oghittu_PRA21}, we use the master equation to derive the differential equations for the expectation value of the squared momentum $\hat{p}_\xi^2$ along the direction $\xi$ (see Appendix~\ref{sec:alternative_derivation} for an alternative derivation):
\begin{equation}
    \begin{split}
        \frac{d}{dt}\langle\hat{p}_\xi^2\rangle=\sum_{\mathbf{q}}\Omega_\mathbf{q}^2q_\xi^2\bigg\{\frac{2\hbar^2}{\omega_\mathbf{q}}\big(2n_\mathbf{q}+1\big)\mathrm{sin}(\omega_\mathbf{q}t)&\\
        +\frac{4\hbar}{M\omega_\mathbf{q}^2}\Big[\omega_\mathbf{q}t\,\mathrm{cos}(\omega_\mathbf{q}t)-\mathrm{sin}(\omega_\mathbf{q}t)\Big]&\langle\hat{p}_\xi^2\rangle\bigg\}
    \end{split}
    \label{eq:p2_equation}
\end{equation}
and for the squared position $\hat{r}_\xi^2$ and covariance $\hat{c}_\xi=\hat{r}_\xi\hat{p}_\xi+\hat{p}_\xi\hat{r}_\xi$:
\begin{equation}
    \begin{split}
        \frac{d}{dt}&\langle\hat{r}_\xi^2\rangle=\frac{1}{M}\langle\hat{c}_\xi\rangle\\[1ex]
        \frac{d}{dt}&\langle\hat{c}_\xi\rangle=\frac{2}{M}\langle\hat{p}_\xi^2\rangle+\\
        &+\sum_\mathbf{q}\Omega_\mathbf{q}q_\xi^2\bigg\{\frac{2\hbar}{M\omega_\mathbf{q}^2}\Big[\omega_\mathbf{q}t\mathrm{cos}\big(\omega_\mathbf{q}t\big)-\mathrm{sin}\big(\omega_\mathbf{q}t\big)\Big]\langle\hat{c}_\xi\rangle\\
        &+\frac{2\hbar^2}{M\omega_\mathbf{q}^2}\big(2n_\mathbf{q}+1\big)\Big[\mathrm{cos}\big(\omega_\mathbf{q}t\big)+\omega_\mathbf{q}t\mathrm{sin}\big(\omega_\mathbf{q}t\big)-1\Big]\bigg\}.
    \end{split}
    \label{eq:r2_c_equations}
\end{equation}
In the limit of a large bath, where $L\to\infty$, the quantized values assumed by the wave vector $q_\xi=2\pi s_\xi/L$ with $s\in\mathbb{Z}$ become closely spaced. In this regime, the sum over $\mathbf{q}$ can be replaced with the integral $L^3/(2\pi)^3\int_{\mathbb{R}^3} d\mathbf{q}$.\\
From the expectation value of $\hat{p}^2=\hat{p}_x^2+\hat{p}_y^2+\hat{p}_z^2$, we calculate the ion temperature. The latter, for an untrapped ion, can be defined as the expectation value of the kinetic energy in units of the Boltzmann constant:
\begin{equation}
    T_\mathrm{ion}=\frac{1}{k_\mathrm{B}}\frac{1}{2M}\langle\hat{p}^2\rangle.
    \label{eq:Tion_def}
\end{equation}
For the sake of completeness, we remark that the definition of the ion temperature can change for different systems. For instance, in the case of Paul-trapped ions, both the secular motion and micromotion have to be taken into account (see Ref.~\cite{Fuerst2018} for details).\\
Finally, we report the equations of motion for the first momenta, which are derived in a similar manner
\begin{equation}
    \begin{split}
        \frac{d}{dt}\langle\hat{r}_\xi\rangle&=\frac{1}{M}\langle\hat{p}_\xi\rangle\\[1ex]
        \frac{d}{dt}\langle\hat{p}_\xi\rangle&=\sum_\mathbf{q}\Omega_\mathbf{q}^2q_\xi^2\bigg\{\frac{2\hbar}{M\omega_\mathbf{q}^2}\Big[\omega_\mathbf{q}t\mathrm{cos}\big(\omega_\mathbf{q}t\big)-\mathrm{sin}\big(\omega_\mathbf{q}t\big)\Big]\langle\hat{p}_\xi\rangle\bigg\}.
    \end{split}
    \label{eq:p_and_x}
\end{equation}

\subsection{Initial \addtwo{quantum} state \addtwo{after ionization}}
\label{sec:initial_state}
\addtwo{The aim of this section is to describe the density matrix of the ion immediately after ionization of a bosonic quantum gas. We assume the BEC with typical parameters presented in Tab.~\ref{tab:experimental_values} is confined in a harmonic potential with trap frequencies $\omega_\xi=2\pi\nu_\xi$, which is typically realized by an optical dipole trap. In the following, we consider two possible experimental scenarios for the ionization process: either ionization of a Rydberg excitation or direct ionization with an ultrashort laser pulse. Let us note that once the ion is created, however, it is no longer affected by the optical dipole trap confining the condensate and no additional external potential for the ion is assumed. Hence, the ion is free to move within the BEC. Nonetheless, the ion inherits its spatial extent, as represented by the squared modulus of its wave function, from the former atom in the trapped condensate before ionization.
Crucially, the spatial extent of the ion must fulfill the requirements of the Lamb-Dicke approximation at all times, as we discuss in Appendix~\ref{app:LD}.} 
Finally, we assume that the ionization process occurs on a time scale much faster than the atomic dynamics, i.e., we treat it as an instantaneous process.\\
\\
\textit{Ionization via Rydberg states -} We begin by considering the case of ionization via Rydberg excitation, for which the initial ionic state can be represented as a thermal state. We assume that, before the ionization, the bosons are in a trapped motional state due to their confinement. At low temperatures, all bosons are described by the same single-particle state, to a very good approximation. If a laser pulse is utilized to excite the internal state of the atoms to a Rydberg state, and if the chosen Rydberg state is such that the corresponding blockade radius is large enough to guarantee a single excitation in the atomic ensemble, then, that excitation is delocalized over the entire atomic cloud. Namely, a giant superposition state is created. The motional state, however, to a very good approximation is the same as before the Rydberg excitation took place. When a second laser pulse is applied to ionize the Rydberg atom as in Ref.~\cite{Engel2018}, the quantum superposition with a single Rydberg excitation is collapsed into a specific product state of the many-body system. Nonetheless, the motional state is still well described by the initial single-particle state of the bosonic ensemble mentioned before, except for an imparted momentum due to the two laser pulses. Specifically, we consider the atom before ionization to be confined in a harmonic trap with trap frequencies $\omega_\xi=2\pi\nu_\xi$ and single-particle eigenenergies $E_{n_\xi}^{(\xi)}=\hbar\omega_\xi n_\xi$ with $n_\xi=0,1,2,\dots$, that is, we neglect interactions among them. Moreover, we assume that the ionization imparts a momentum $k_{0,\xi}$ along the $\xi$-direction at $t=t_0$. Assuming that the atom is not completely cooled down to the trap ground state, the density matrix reads
\begin{equation}
    \begin{split}
        \hat{\rho}_\xi(t_0)=\Big(1-&e^{-\frac{\hbar\omega_\xi}{k_\mathrm{B}T_\mathrm{gas}}}\Big)e^{ik_{0,\xi}\hat{r}_\xi}\times\\
        &\times\sum_{n_\xi} e^{-\frac{E_{n_\xi}^{(\xi)}}{k_\mathrm{B}T_\mathrm{gas}}}\ket{n_\xi}\bra{n_\xi}e^{-ik_{0,\xi}\hat{r}_\xi},
    \end{split}
    \label{eq:rho_xi}
\end{equation}
where $\ket{n_\xi}$ are the states of the harmonic oscillator with frequency $\omega_\xi$ and $T_\mathrm{gas}$ the gas temperature.\\
The initial value of the squared momentum along the direction $\xi$ is calculated as the average $\mathrm{Tr}\{\hat{p}_\xi^2\hat{\rho}_\xi(t_0)\}$ over the initial density matrix $\hat{\rho}_\xi(t_0)$. Using the definition of the momentum operator $\hat{p}_\xi=i\sqrt{\hbar M \omega_\xi/2}(\hat{a}^\dag-\hat{a})$ and the properties of the trace, we get to the following formula for the initial squared momentum \footnote{In the derivation of Eq.~\eqref{eq:p02_formula} we use the \addtwo{Baker-Campbell-Hausdorff (BCH) identity} $e^{i\hat{G}\lambda}\hat{A}e^{-i\hat{G}\lambda}=\hat{A}+i\lambda\comm{\hat{G}}{\hat{A}}+\frac{(i\lambda)^2}{2!}\comm{\hat{G}}{\comm{\hat{G}}{\hat{A}}}+\dots$, whose terms of order higher than two are zero for $\hat{G}=\hat{r}_\xi$ and $\hat{A}=\hat{p}^2$.}:
\begin{equation}
    \begin{split}
        \langle\hat{p}_{\xi}^2(t_0)\rangle=&\Big(1-e^{-\frac{\hbar\omega_\xi}{k_\mathrm{B}T_\mathrm{gas}}}\Big)\times\\
        &\times\sum_n e^{-\frac{\hbar\omega_\xi}{k_\mathrm{B}T_\mathrm{gas}}n}\bigg[\frac{\hbar M \omega_\xi}{2}\big(2n+1\big)+\hbar^2k_{0,\xi}^2\bigg].
    \end{split}
    \label{eq:p02_formula}
\end{equation}
In a similar fashion, we obtain the initial squared position:
\begin{equation}
    \langle\hat{r}_{\xi}^2(t_0)\rangle=\Big(1-e^{-\frac{\hbar\omega_\xi}{k_\mathrm{B}T_\mathrm{gas}}}\Big)\sum_n e^{-\frac{\hbar\omega_\xi}{k_\mathrm{B}T_\mathrm{gas}}n}\frac{\hbar}{2M\omega_\xi}\big(2n+1\big),
    \label{eq:x02_formula}
\end{equation}
whereas the initial values of the covariance $\hat{c}_\xi$ is always zero.\\
\\
\begin{table}[ht]
    \centering
    \setlength{\extrarowheight}{1ex}
    \setlength{\tabcolsep}{2ex}
    \begin{tabular}{l c}
        \hline
        \multicolumn{2}{l}{\textbf{Ion}} \\
        \hline
        Kinetic energy & $1.3\cdot10^{-7}\,\mathrm{eV}$\\
        Temperature & $1\,\mathrm{mK}$\\
        Excess velocity & $530\,\mathrm{mm}/\mathrm{s}$\\[2ex]
        \hline\hline
        \multicolumn{2}{l}{\textbf{BEC}} \\
        \hline
        Atom number & $3\cdot10^{4}$\\
        Peak density & $2\cdot10^{14}\,\mathrm{cm}^{-3}$\\
        Speed of sound & $2.7\,\mathrm{mm}/\mathrm{s}$\\
        Trap frequencies $\nu_\xi$ & $120-170\,\mathrm{Hz}$\\
        Cloud radius & $5\,\mathrm{\mu m}$\\
        \hline
    \end{tabular}
    \caption{Typical experimental parameters for the homonuclear system $^{87}$Rb$^+$/$^{87}$Rb. The initial kinetic energy of the $^{87}$Rb$^+$ ion corresponds to a two-photon ionization via a virtual intermediate state by an intense femtosecond laser pulse with a duration of $200\,\mathrm{fs}$ near the ionization threshold. The parameters of the Bose-Einstein condensate are typical for $^{87}$Rb atoms in an optical dipole trap.}
    \label{tab:experimental_values}
\end{table}
\\
\textit{Ionization with an ultrashort laser pulse -} Another interesting scenario is the ionization procedure employed in Ref.~\cite{Simonet_NC21}, where a femtosecond laser is focused down to a waist $w_0=1\,\mathrm{\mu m}$, which is small compared to the size of the atomic cloud. Within a single pulse of $215\,\mathrm{fs}$ duration, the number of ionized atoms can be precisely tuned with the laser peak intensity. More details of the experimental procedure are reported in Sec.~\ref{sec:exp}. In this case, the ionization process can be interpreted as a continuous measurement process, where the focused laser beam with Gaussian envelop $e^{-2\mathbf{r}^2/w_0^2}$ is the probe field~\cite{Steck_CP06}. Therefore, the probability of finding the ion at position $\mathbf{r}$ is given by
\begin{equation}
    P(\mathbf{r})=\sqrt{\frac{2}{\pi w_0^2}}\int_{\mathbb{R}^3}d\mathbf{r}'\,e^{-\frac{2}{w_0^2}(\mathbf{r}'-\mathbf{r})^2}\bra{\mathbf{r}'}\hat{\rho}_\mathrm{BEC}\ket{\mathbf{r}'}
    \label{eq:probability}
\end{equation}
namely, the convolution between the Gaussian beam and the probability density $\bra{\mathbf{r}'}\hat{\rho}_\mathrm{BEC}\ket{\mathbf{r}'}$ of the condensate. The initial density distribution of the ion can be therefore identified with Eq.~\eqref{eq:probability}, while the initial ion's wave function can be defined, apart from a global phase, as the square root of $P(\mathbf{r})$, with spatial extent determined by the beam-waist. 
Consider an ultra-cold bosonic gas with experimental parameters as listed in Tab.~\ref{tab:experimental_values} that corresponds to the experimental situation reported in Ref.~\cite{Simonet_NC21}. Hence, the bosonic density distribution is well described by the Thomas-Fermi profile, which reads
\begin{equation}
    \bra{\mathbf{r}'}\hat{\rho}_\mathrm{BEC}\ket{\mathbf{r}'}=n_0\Bigg[1-\bigg(\frac{x'}{R_x}\bigg)^2-\bigg(\frac{y'}{R_y}\bigg)^2-\bigg(\frac{z'}{R_z}\bigg)^2\Bigg]
    \label{eq:TFprofile}
\end{equation}
in the region defined by the ellipsoid with radii $R_\xi$ ($\xi=x,y,z$), and zero elsewhere. The definitions of $R_\xi$ and other details on the Thomas-Fermi approximation can be found, e.g., in Ref.~\cite{PitaevskiiStringari}. The integral in Eq.~\eqref{eq:probability} can be computed numerically in spherical coordinates. As anticipated, we define the initial ion wavefunction as $\Psi_0(\mathbf{r})=e^{ik_{0,\xi}\hat{r}_\xi}\sqrt{P(\mathbf{r})}$, where we added the contribution of the initial imparted momentum along the $\xi$ direction.\\
\\
The initial state of the ion is used to calculate the initial conditions for the equations of motion of the expectation values given in Sec.~\ref{sec:LDapprox}. For the parameters considered in our study, however, no significant differences have been observed between the initial states obtained after photoionization of a Rydberg atom or of a ground-state atom with a femtosecond laser pulse. Albeit the numerical analysis in the following section refers to the thermal state~(\ref{eq:rho_xi}), we note that the choice of one or the other initial condition does not affect the conclusions we are going to outline in Sec.~\ref{sec:results}.

\section{Results}
\label{sec:results}
In this section, we report on the dynamics of an ion with initial momentum in an ultracold bosonic cloud. The evolution of the ion temperature, velocity and position are obtained by numerically solving Eq.~\eqref{eq:p2_equation} and Eq.~\eqref{eq:p_and_x}. We investigate the impact of different experimental parameters such as the initial momentum of the ion $k_0$, the density of the atomic cloud $n_0$ and the atom-ion scattering length $a_\mathrm{ai}$ on the ion dynamics. Unless stated differently, the system consists of a $^{87}\mathrm{Rb}^+$ ion in a bosonic bath of $^{87}\mathrm{Rb}$ atoms at $T_\mathrm{gas}=1\,\mathrm{nK}$, with $n_0=2\cdot10^{14}\,\mathrm{cm}^{-3}$, and $a_\mathrm{ai}\simeq R^\star$ corresponding to the potential in Fig.~\ref{fig:potential} (see also Tab.~\ref{tab:experimental_values}). Let us note that the results obtained at fixed density do not depend on the specific value of the temperature of the ultracold bosonic gas, $T_\mathrm{gas}$, which is chosen according to the discussion in Appendix~\ref{app:LD}. Moreover, we consider the momentum imparted at $t=t_0$ to be directed along $x$, and we focus on the dynamics along the same direction. In fact, although the initial conditions for $T_\mathrm{ion}^{y,z}$ may be different from zero depending on the choice of the initial state and the direction of the imparted momentum, the decoupling of the three directions allows us to consider just one direction without any loss of generality \footnote{A finite initial temperature along $y$ and $z$ would not have any qualitative effect on our analysis. However, a quantitative study could require the total temperature $T_\mathrm{ion}=T_\mathrm{ion}^x+T_\mathrm{ion}^y+T_\mathrm{ion}^z$ to be considered.}.\\
\\
\begin{figure}
    \centering
    \includegraphics[width=.45\textwidth]{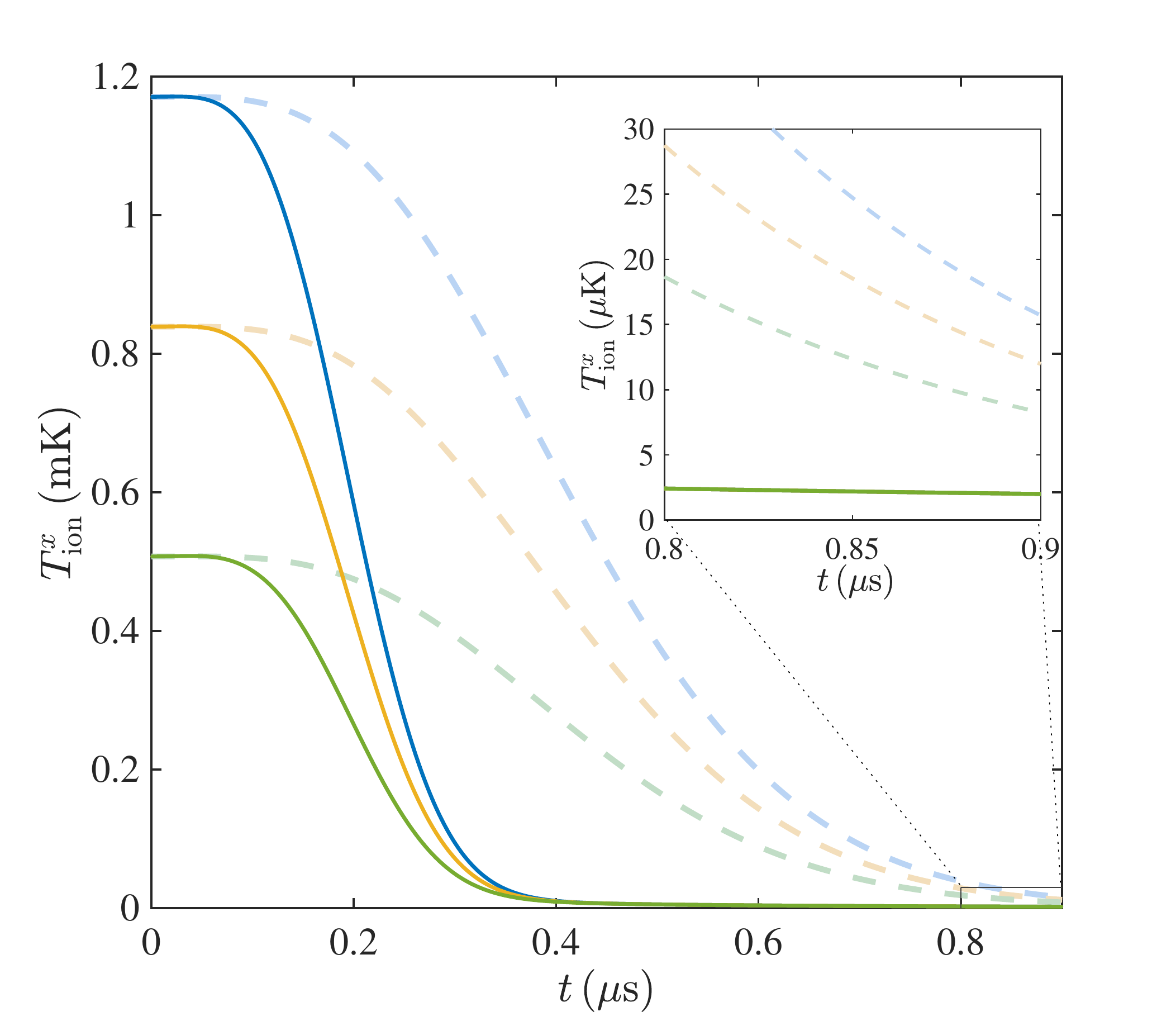}
    \caption{Ion temperature $T_\mathrm{ion}^x=\langle\hat{p}_x^2\rangle/(2Mk_\mathrm{B})$ as a function of time for $n_0=2\cdot10^{14}\,\mathrm{cm}^{-3}$ (solid lines) and $n_0=2\cdot10^{13}\,\mathrm{cm}^{-3}$ (light dashed lines). \textbf{(a)} The initial ion temperatures correspond to $T_\mathrm{ion}^x=1.17\,\mathrm{mK}$ (blue), $T_\mathrm{ion}^x=0.84\,\mathrm{mK}$ (orange) and $T_\mathrm{ion}^x=0.51\,\mathrm{mK}$ (green). The inset shows a magnification of the main plot in the range from $t=0.8\,\mu\mathrm{s}$ to $t=0.9\,\mu\mathrm{s}$, as indicated. From the latter, we observe that the temperature corresponding to $n_0=2\cdot10^{14}\,\mathrm{cm}^{-3}$ converges to a value of around $2\,\mathrm{\mu K}$, independent on the initial condition.}
    \label{fig:Ttot}
\end{figure}
\begin{figure}
    \centering
    \includegraphics[width=.45\textwidth]{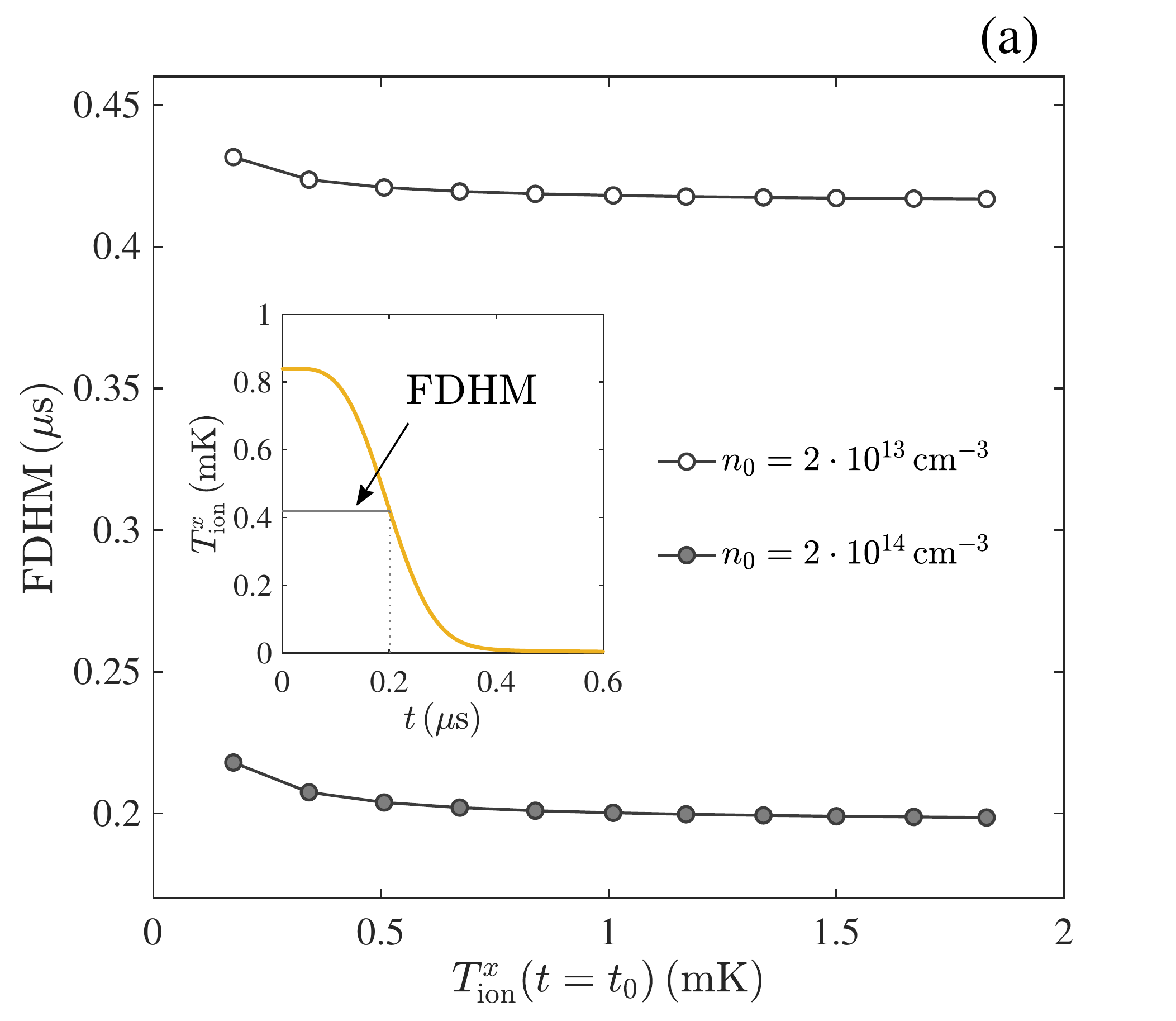}\\
    \includegraphics[width=.45\textwidth]{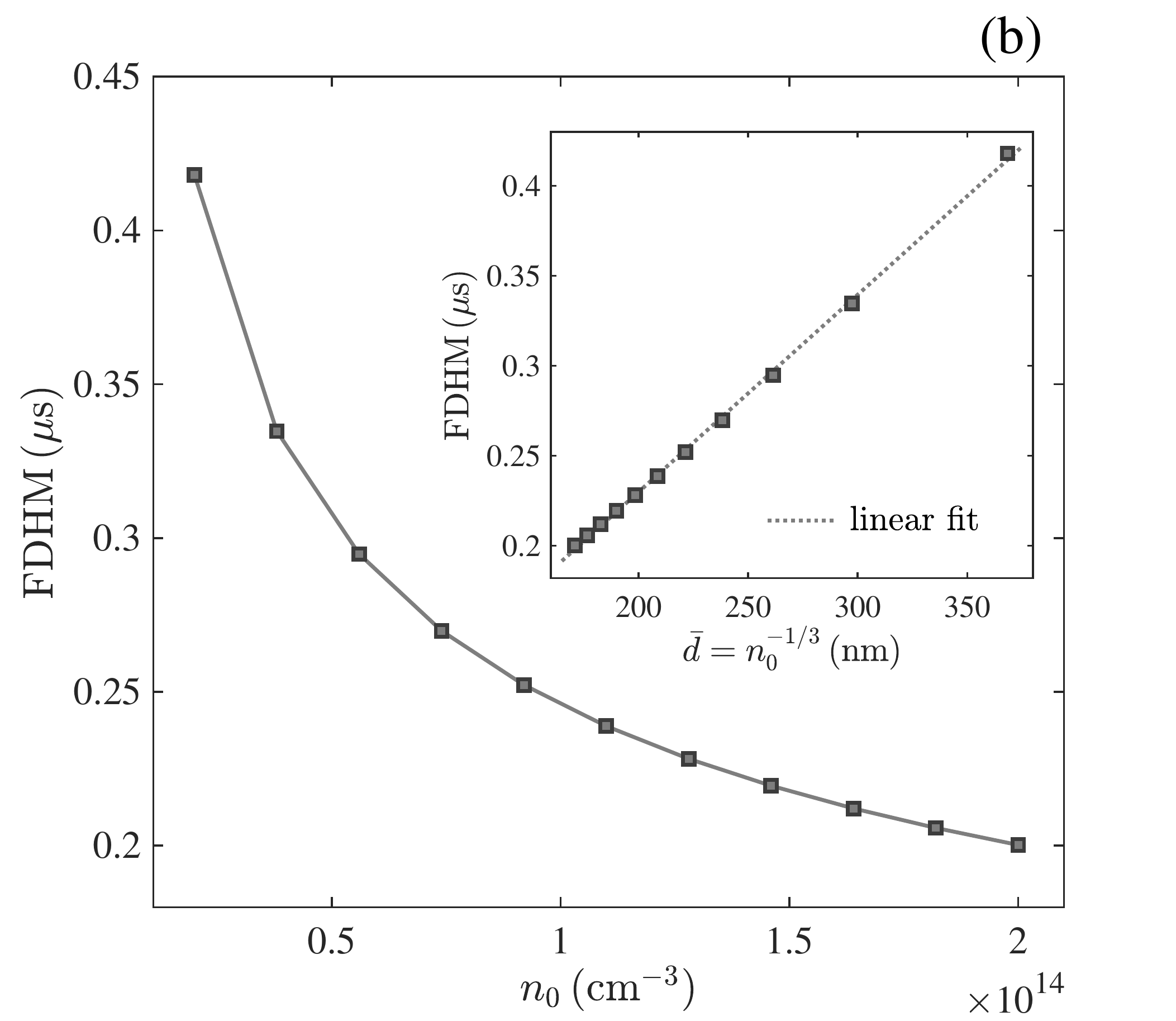}
    \caption{\textbf{(a)} Main plot: full duration at half maximum (FDHM) for two atomic densities as a function of the initial ion temperature; solid lines connecting the points are a mere guide to the eye. Inset: definition of FDHM. \textbf{(b)} FDHM for $T_\mathrm{ion}^x(t=t_0)\simeq1.01\,\mathrm{mK}$ and $a_\mathrm{ai}\simeq R^\star$ (gray squares). Main plot: FDHM as a function of the gas density $n_0$; the solid line connecting the points is a mere guide to the eye. Inset: FDHM as a function of the average particle separation $\Bar{d}=1/\sqrt[3]{n_0}$; the dotted line is a linear function fitting the data.}
    \label{fig:WHM}
\end{figure}
\\
\subsection{Cooling dynamics}
We start by comparing the ion temperature as a function of time for different initial conditions.\\
\\
\textit{Initial temperature of the ion -} In Fig.~\ref{fig:Ttot}, the results corresponding to initial ion temperatures in the millikelvin regime are shown. We can observe from the main plot and inset that the time required for $T_\mathrm{ion}^x$ to converge to $2\,\mathrm{\mu K}$ is almost independent on its initial value at $t=t_0$. In other words, the larger the initial momentum, the faster the cooling. On the other hand, the cooling dynamics is strongly affected by the condensate density $n_0$, as can be observed by comparing the dark solid lines with the light dashed lines. \addone{We refer to Appendix~\ref{app:low_T} for a comparison with the dynamics corresponding to lower initial ion temperatures.}\\
\\
\textit{Atomic density -} To systematically study the cooling dynamics, we define the full duration at half maximum (FDHM) as the time it takes for the ion temperature to reach half of its initial value [see inset of Fig.~\ref{fig:WHM}(a)]. Note that small values of the FDHM correspond to higher cooling rates: \addone{the larger is the FDHM, the smaller is the atom-ion cross section and vice versa.} The time-scale of the cooling dynamics is similar to the average time-scale for classical collisions with one atom in the bath given by $t = d_\mathrm{WS}/v_x(t_0) = 265\,\mathrm{ns}$ for an initial velocity of $v_x(t_0)=0.4\,\mathrm{m}\,\mathrm{s}^{-1}$ and with $d_\mathrm{WS} = \left( 3/(4 \pi n_0) \right)^{1/3} = 106\,\mathrm{nm}$ being the Wigner-Seitz radius at a condensate density of $n_0=2\cdot10^{14}\,\mathrm{cm}^{-3}$.\\
The circles in the main plot of Fig.~\ref{fig:WHM}(a) show that the FDHM is barely affected by the initial temperature of the ion. Moreover, the same weak dependence is observed for $n_0=2\cdot10^{14}\,\mathrm{cm}^{-3}$ (full circles) and $n_0=2\cdot10^{13}\,\mathrm{cm}^{-3}$ (empty circles). Figure~\ref{fig:WHM}(b) quantifies how effective the cooling of the ion is, depending on the density of the condensate (main plot) and on the mean distance between the atoms (inset). The initial temperature is fixed to $T_\mathrm{ion}^x(t_0)\simeq 1.1\,\mathrm{mK}$ and the gray squares are the values of the FDHM for different $n_0$ (or $\Bar{d}=1/\sqrt[3]{n_0}$ in the inset). As expected, a denser gas ensures a faster cooling (i.e. a smaller FDHM) due to the stronger impact of the atom-ion interaction on the ion dynamics.
We observe that the FDHM increases linearly with the mean distance. Both the gas and the ion are treated fully three-dimensionally in the master equation. However, due to the Lamb-Dicke approximation, solutions are given by the tensor product of the density matrices of the three spatial directions. Fig.~\ref{fig:WHM}(b) exemplary shows the result for the $x$ direction as the dynamics is effectively one-dimensional for the ion moving into a fixed direction. 
Because of this, the ion dynamics is characterized by the mean distance between the bosons, which accounts for the rate of atom-ion collisions in one direction: the larger the distance, the larger the FDHM, i.e., the smaller is the cooling rate and vice versa. This is in contrast with the expectation that the cooling rate is linearly proportional to the gas density. In the future it will be interesting to find solutions to solve the master equation without relying on the Lamb-Dicke approximation to investigate the density dependence of the FDHM as well as a maximum capture velocity for the cooling and pinning dynamics.\\
\begin{figure}
    \centering
    \includegraphics[width=.45\textwidth]{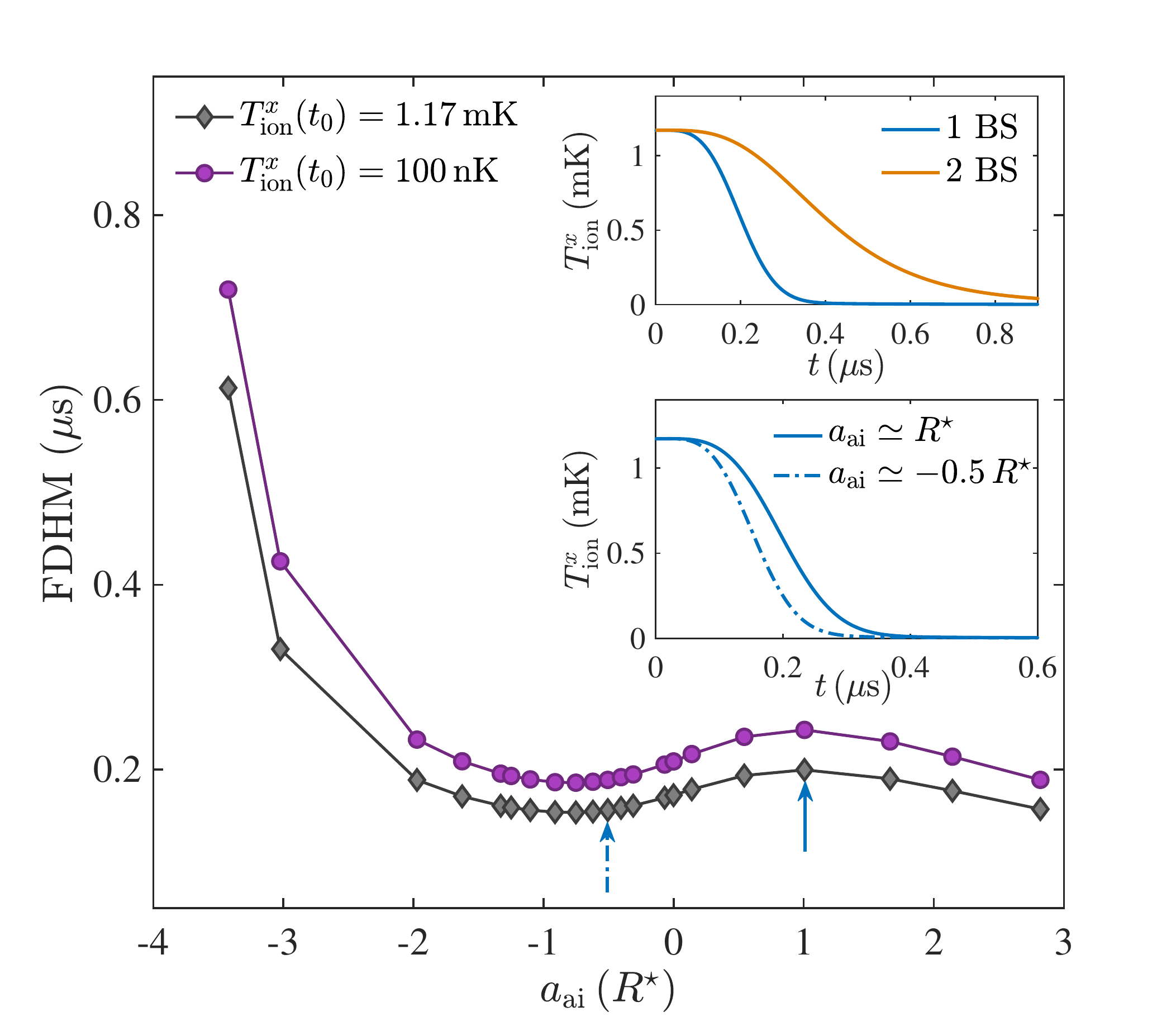}
    \caption{\addone{Main plot: scattering length dependence of the FDHM for $T_\mathrm{ion}^x(t_0)=1.17\,\mathrm{mK}$ (gray diamonds) and $T_\mathrm{ion}^x(t_0)=100\,\mathrm{nK}$ (purple circles). The lines are a mere guide to the eye. Lower inset: time dependence of the ion temperature for two values of the atom-ion scattering length and initial ion temperature $T_\mathrm{ion}^x=1.17\,\mathrm{mK}$. The two values of the scattering length are indicated in the main plot by the two arrows. Upper inset: time dependence of the ion temperature for two regularized potentials supporting a different number of two-body bound states: one bound state with $E_\mathrm{BS}\simeq-1.43\,E^\star$ (blue) and two bound states with $E_\mathrm{BS,1}\simeq-1.39\,E^\star$ and $E_\mathrm{BS,2}\simeq-152.78\,E^\star$ (orange).}}
    \label{fig:p2scattering}
\end{figure}
\\
\textit{Atom-ion scattering length -} Another feature that we point out is the dependence of the cooling dynamics on the atom-ion scattering length. The recent observation of Feshbach resonances in compound atom-ion systems~\cite{Schaetz_NAT21} confirms the possibility of tuning the atom-ion interaction via an external magnetic field. This dependence can be exploited in experiments to achieve a higher cooling rate without changing parameters such as the atomic density or the ion initial temperature.\\
\addone{Since the cooling dynamics is closely related to the elastic cross-section, no strong dependence on the scattering length would be expected at high collision energies, where the ion can be treated classically. In contrast, such a dependence could be expected at ion temperatures on the order of $\mu$K and below, where fewer partial waves contribute to scattering events and quantum effects become relevant. On this regard, we note that the number of partial waves contributing in the millikelvin regime is on the order of ten.
In the main plot of Fig.~\ref{fig:p2scattering} the non-trivial dependence of the FDHM on $a_\mathrm{ai}$ is shown for two values of the initial temperature: $T_\mathrm{ion}^x(t_0)=1.17\,\mathrm{mK}$ and $T_\mathrm{ion}^x(t_0)=100\,\mathrm{nK}$, the latter being on the order of the typical energy of the atom-ion potential $E^\star/k_B = 79\,\mathrm{nK}$.
We observe a similar behavior for the two initial conditions. 
However, considering values of $a_\mathrm{ai}$ between $\sim-2\,R^\star$ and $\sim2\,R^\star$, the difference between the maximum and minimum FDHM for the lower initial temperature of the ion is $24\%$ larger compared to the higher initial temperature ($0.057\,\mathrm{\mu s}$ and $0.046\,\mathrm{\mu s}$, respectively). This shows that the dependence is indeed more pronounced when the collision energy is lower, as expected. The noticeably larger values of the FDHM observed for the two values of $a_\mathrm{ai}$ below $-2\,R^\star$ could indicate the failure of the Born approximation due to the strong atom-ion coupling. Another hypothesis to explain the dependence of the FDHM on $a_\mathrm{ai}$ could be the binding of atoms to the ion. This would increase the effective mass of the ion, which would modify the scattering parameters with the bath as well as the cooling dynamics.\\
Finally, we consider a regularized atom-ion potential supporting two two-body bound states. In the upper inset of Fig.~\ref{fig:p2scattering} we can observe that the cooling dynamics does not depend qualitatively on the number of such bound states. Although the FDHM corresponding to two bound states is about twice the value obtained with one bound state, the reduction of $T_\mathrm{ion}^x$ takes place on similar timescales in the two cases. We remark, however, that the choice of the potential with one bound state is justified by the fact that the occupation of deeply bound states is much less likely compared to the occupation of loosely bound states because of the large energy gap between them (see, e.g. \cite{CotePRL02}).}\\
\begin{figure}
    \centering
    \includegraphics[width=.45\textwidth]{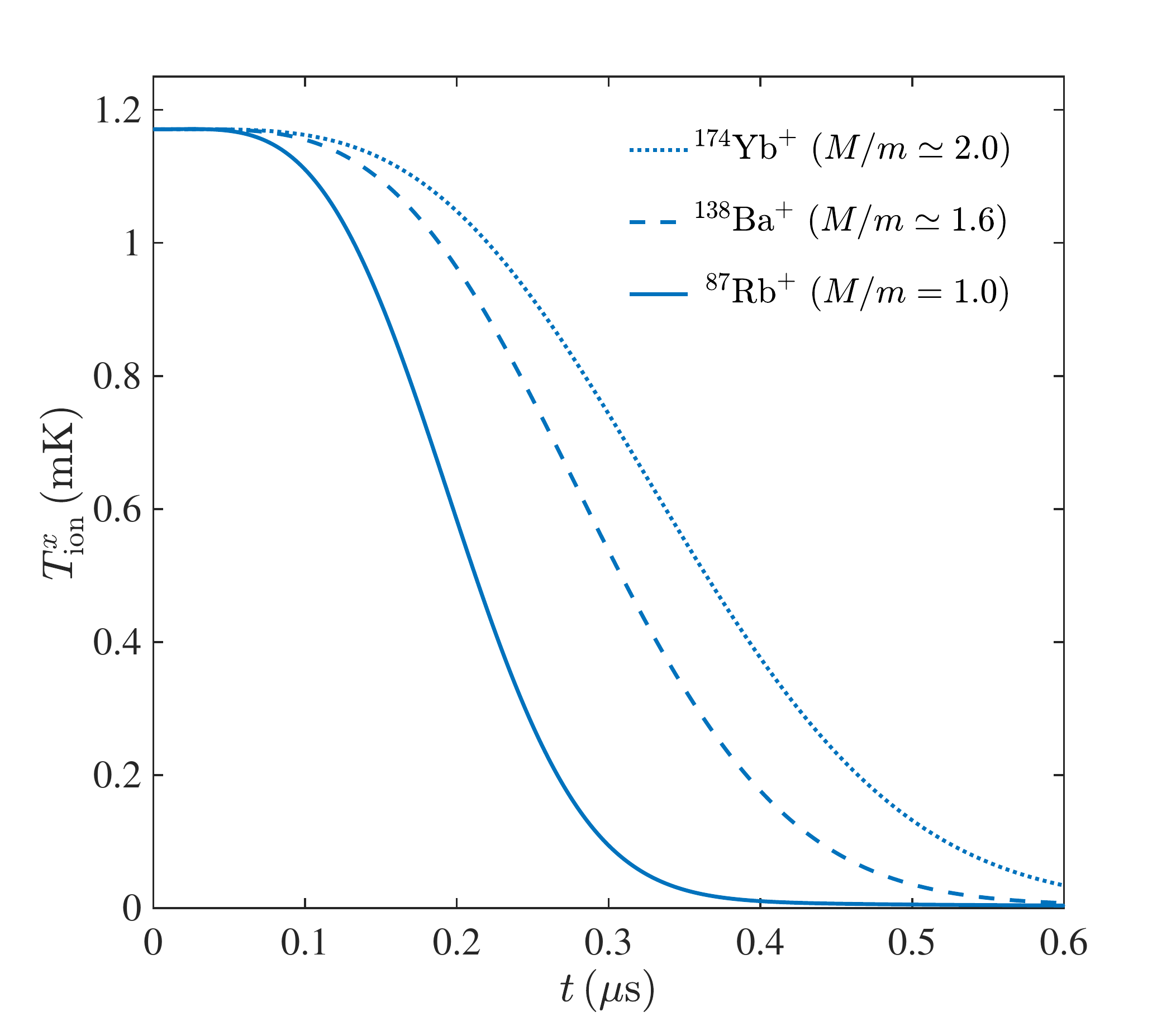}
    \caption{Time dependence of ion temperature for different ion species immersed in a gas of $^{87}$Rb atoms with density $n_0=2\cdot10^{14}\,\mathrm{cm}^{-3}$. The initial ion temperature is $T_\mathrm{ion}^x=1.17\,\mathrm{mK}$. The corresponding values of $R^\star$ are the following: $265.81\,\mathrm{nm}$ for $^{87}$Rb$^+$, $294.67\,\mathrm{nm}$ for $^{138}$Ba$^+$ and $307.23\,\mathrm{nm}$ for $^{174}$Yb$^+$.}
    \label{fig:T_different_ions}
\end{figure}
\textit{Ionic species -} Similar simulations are repeated for different ionic species moving in the $^{87}$Rb atomic gas. In Fig.~\ref{fig:T_different_ions}, the time dependence of $T_\mathrm{ion}^x$ is shown for $^{138}\mathrm{Ba}^+$ and $^{174}\mathrm{Yb}^+$ compared to the rubidium ion considered in the previous analysis. The observed behavior is qualitatively the same, but the plot shows that higher values of the ion-atom mass ratio $M/m$ result in slower cooling. We remark that the use of different ions affects the value of the ratio $M/m$, hence, the range of validity of the Lamb-Dicke approximation (see Appendix~\ref{app:LD}).\\
\\
\subsection{Pinning dynamics} 
\begin{figure}
    \centering
    \includegraphics[width=.45\textwidth]{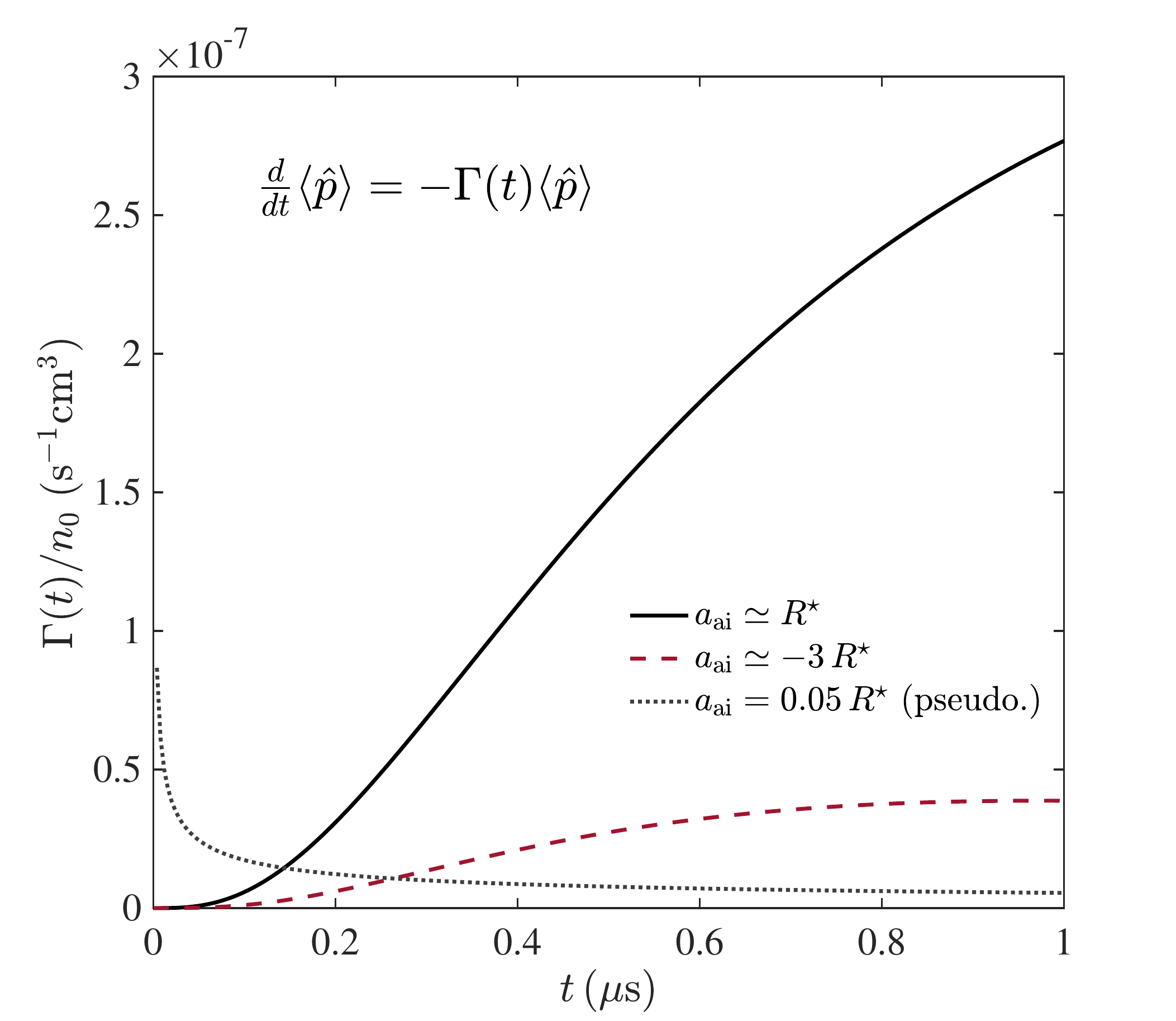}
    \caption{Time-dependent friction in units of the condensate density $n_0$ as a function of time for two different atom-ion scattering lengths \addone{(solid and dashed) and for a neutral impurity (dotted)}.}
    \label{fig:friction}
\end{figure}

Now, we discuss the evolution of the expectation value of the position and momentum of the ion given by Eq.~\eqref{eq:p_and_x}. Note that all results are given in one dimension, since the initial momentum $k_0$ is assumed to be along $x$.\\
\begin{figure}
    \centering
    \includegraphics[width=.45\textwidth]{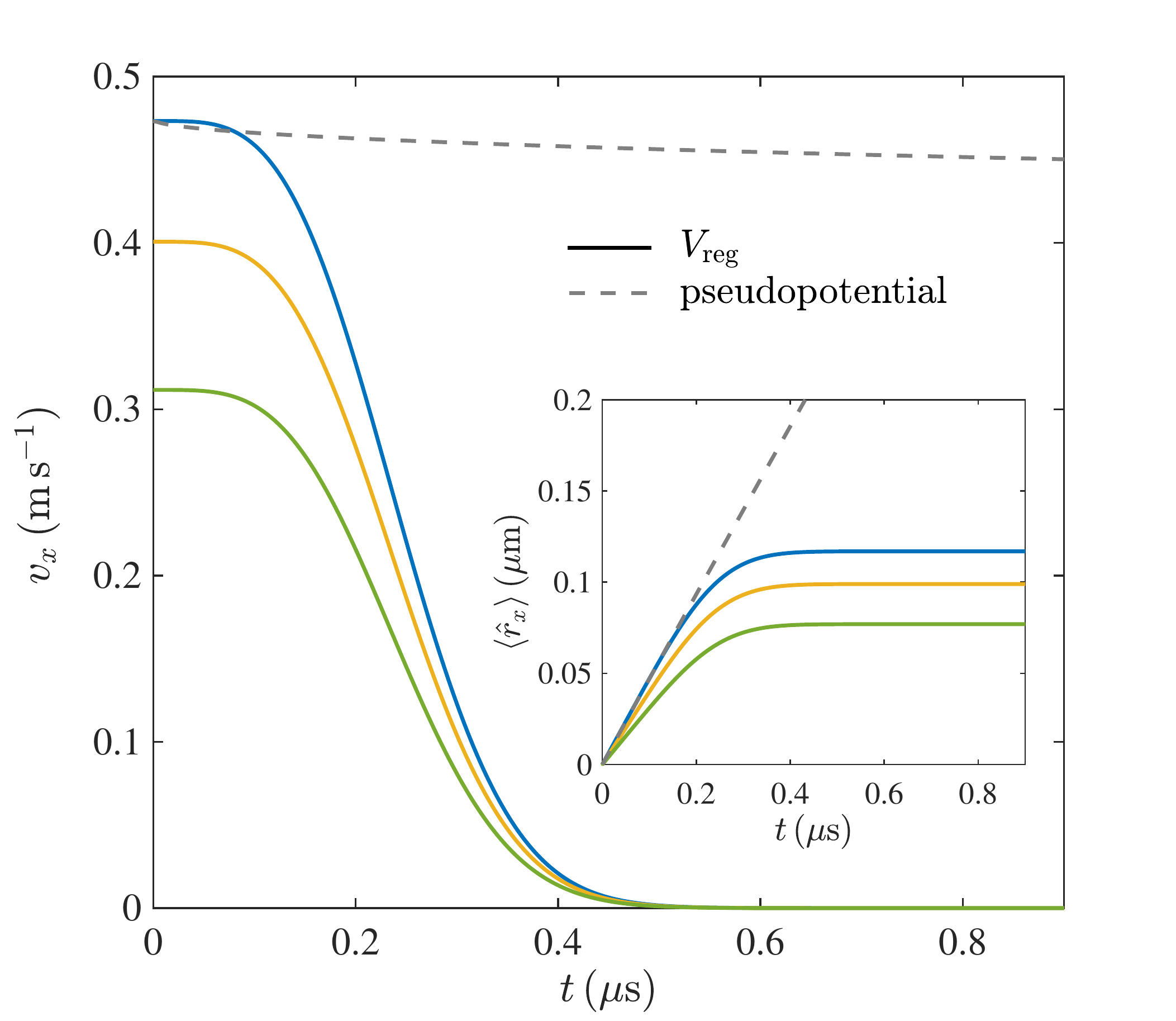}
    \caption{Time dependence of the ion's velocity $v_x=\langle\hat{p}_x\rangle/M$ (main plot) and position (inset) along the $x$-direction for different initial ion velocities: $v_x=0.47\,\mathrm{m\,s^{-1}}$ (blue), $v_x=0.40\,\mathrm{m\,s^{-1}}$ (orange) and $v_x=0.31\,\mathrm{m\,s^{-1}}$ (green). Solid lines correspond to a regularized atom-ion potential with $a_\mathrm{ai}\simeq R^\star$, whereas the gray dashed line corresponds to a neutral impurity in a gas with a short-range pseudopotential with $a_\mathrm{ai}\simeq0.05\,R^\star$.}
    \label{fig:first_moments}
\end{figure}
\begin{figure}
    \centering
    \includegraphics[width=.45\textwidth]{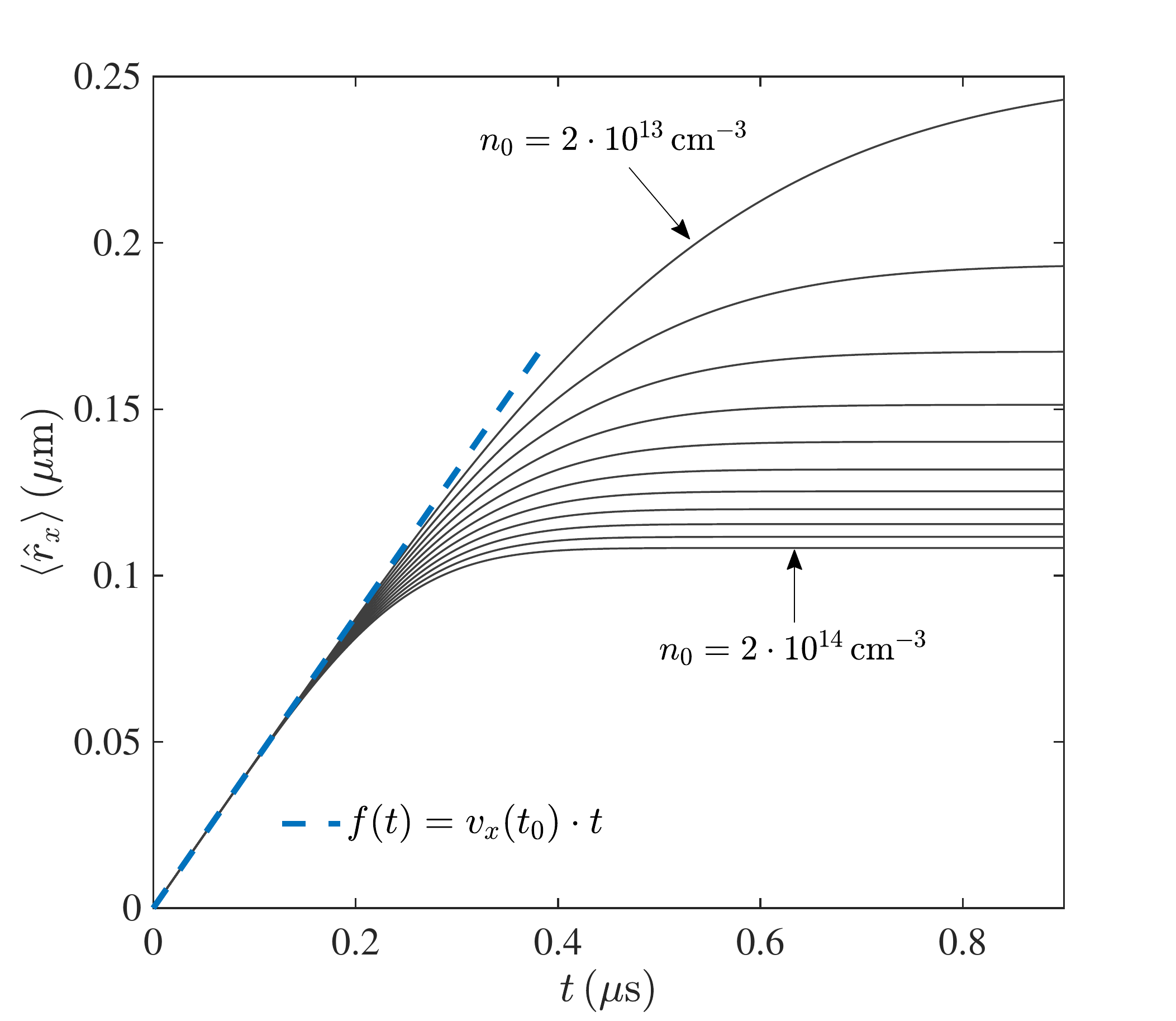}
    \caption{Time dependence of the ion's position for $T_\mathrm{ion}^x(t_0)=1.17\,\mathrm{mK}$ and different gas densities (solid gray lines). The values of $n_0$ are chosen uniformly in the interval between the indicated values of $n_0=2\cdot10^{13}\,\mathrm{cm}^{-3}$ and $n_0=2\cdot10^{14}\,\mathrm{cm}^{-3}$. The dashed blue line represents the position of a particle moving at constant velocity.}
    \label{fig:rx_density}
\end{figure}
\\
\textit{Ion velocity evolution -} In the main plot of Fig.~\ref{fig:first_moments}, we observe how the ion's velocity decays, reaching a value on the order of $10^{-10}\,\mathrm{m}\,\mathrm{s}^{-1}$ at $t=0.9\,\mathrm{\mu s}$. Similar to what was observed for the decay of $T_\mathrm{ion}$, the time required for the velocity to decay depends only weakly on its initial value. For this reason, the ion's final positions are reached at approximately the same time for all values of $v_x(t_0)=\hbar k_0/M$, as shown in the inset of Fig.~\ref{fig:first_moments}.\\
\addtwo{This result is completely different from the dynamics of a neutral impurity in a bosonic bath interacting via a short-range pseudopotential, as shown in Fig.~\ref{fig:first_moments} by the dashed gray lines. On the timescale relevant for the atom-ion dynamics, the neutral impurity does not come to rest, and the neutral impurity moves with constant velocity through the gas (see inset). We attribute this difference to the long-range character of the polarization potential, which cannot be adequately described by taking only the $s$-wave scattering into account.} For a meaningful comparison, we choose a value of $0.05\,R^\star$ for the impurity-gas scattering length, corresponding roughly to the range of the van der Waals interaction between $^{87}$Rb atoms. \addtwo{Note that choosing a scattering length comparable to $R^\star$ for the neutral impurity would correspond to the unitary limit, and the validity of the master equation description would likely no longer hold.}\\
\\
\textit{Ion position evolution -} The onset of the pinning dynamics is affected by the gas density as shown in Fig.~\ref{fig:rx_density}. There, the dashed blue line represents the position of a particle moving with constant velocity, while the gray solid lines correspond to the ion's position in the presence of a condensate with different densities. The plot shows that, at short times, the ion's position is not affected by the presence of the gas, while at later times it is deflected to its final value at a rate increasing with the density.\\
Interestingly, the initial linear time dependence of the ion dynamics in the gas is an indication of the polaronic behavior. Specifically, due to its interaction with the bosonic bath, the ion is dressed by phononic excitations in such a way that it can be considered as a quasi particle moving freely within the gas. However, as time evolves, effects such as dephasing of the phonon modes become dominant until the ion comes to rest.\\
\\
\textit{Friction coefficient evolution -} Since the motion of the ion cannot be explained by a classical trajectory, we have analyzed the equation more closely for the expectation value of the momentum [second line of Eq.~\eqref{eq:p_and_x}]. That equation can be compared to the classical equation of a particle subject to a friction. On this purpose, we rewrite it as
\begin{equation}
    \frac{d}{dt}\langle\hat{p}_\xi\rangle=-\Gamma(t)\langle\hat{p}_\xi\rangle,
\end{equation}
where we defined the friction $\Gamma(t)$ according to Eq.~\eqref{eq:p_and_x}.\\
In Fig.~\ref{fig:friction} the time dependence of $\Gamma$ for two values of the scattering length \addone{and for a neutral particle} is shown. In a classical scenario the friction coefficient would be constant in time, whereas here it is explicitly time-dependent. Since all properties of the atomic bath are constant in time, the time-dependent friction observed here can only be explained by a change in the properties of the impurity. \addone{Moreover, we note that the qualitative difference between the friction coefficient corresponding to the neutral and charged particle highlights the key role of the long-range atom-ion potential in our predictions.
In the neutral case, the time evolution of the friction can be associated to the formation of a Bose polaron: as the impurity moves through the condensate, it gets dressed by phononic excitations, resulting in the reduction of the friction coefficient.} For the ionic impurity, a transient phase is observed at very short times where the friction is almost zero for both scattering lengths. This phase corresponds to the regime where the particle is not affected by the presence of the gas, as shown in Fig.~\ref{fig:rx_density}, that is, a polaronic-type behavior. The increase in $\Gamma$ at longer times is responsible for the pinning dynamics. We note that for large negative scattering lengths, we observe much smaller values of $\Gamma$, corresponding to a slower pinning. While for shorter time-scales the Bogoliubov phonon modes behave coherently due to the superfluidity of the bath, at longer times coherence is reduced, which we attribute to dephasing of the phononic modes.
Whether this phenomenon is connected to the formation of two-body atom-ion states supported by our regularized interaction is not possible to quantify in the current formulation of the master equation. It will be interesting in the future to investigate whether the formation of many-body bound states as those predicted in Refs.~\cite{Gregory2021,Bruun_PRL21,SchurerPRL17} occur and to understand if they are responsible of the pinning dynamics we observe in this work. For this, the description of the atomic gas needs to be modified and the back action of the ion on the atomic gas has to be included.

\section{Experimental considerations}
\label{sec:exp}
In this section, we will describe experimental settings that should allow investigating the cooling and pinning dynamics of an ion in an ultracold bosonic gas. Finally, previously neglected inelastic processes and their impacts on the ionic dynamics are discussed.\\
\\
\textit{Results validation in future experiments -} In order to experimentally validate the calculated dynamics, it is necessary to instantaneously create an ion out of an ultracold bosonic gas with a defined, but tunable, initial velocity $v_x$. Two-photon ionization via a virtual intermediate state by an intense femtosecond laser pulse with adjustable wavelength gives rise to a tunable excess energy $E_\mathrm{exc}$. This results in an adjustable initial velocity of the ion (compare Fig.~\ref{fig:scheme}). The experiments in Refs.~\cite{Wessels_CP18,Simonet_NC21} use ultrashort laser pulses of $\sim200\,\mathrm{fs}$ duration and a rather high excess energy of $E_\mathrm{exc}=0.68\,\mathrm{eV}$, which corresponds to an initial kinetic energy of the $^{87}$Rb ion of $\sim4\,\mathrm{\mu eV}$ or an initial temperature of $T_\mathrm{ion}=33\,\mathrm{mK}$ \cite{Simonet_NC21}. However, by using an optical-parametric amplifier, the wavelength of the laser pulses can be tuned close to the ionization threshold so that the excess energy is ultimately limited by the bandwidth of the laser pulse due to the time-energy uncertainty relation. A Gaussian laser pulse of $200\,\mathrm{fs}$ duration corresponds to a kinetic energy of $115\,\mathrm{neV}$ for a $^{87}$Rb ion, which relates to an initial temperature of $T_\mathrm{ion}=890\,\mathrm{\mu K}$. Such a regime is covered by the initial parameters of our calculations and would allow stopping the ion within the BEC.\\
In order to resolve the cooling and pinning dynamics of the ion, it needs to be created in a localized region much smaller than the extent of the BEC. This is possible by focusing the laser beam to a diffraction limited spot with a high-resolution microscope objective \cite{Simonet_NC21}. Because two $594\,\mathrm{nm}$ photons are sufficient to excite the outermost electron of $^{87}$Rb just over the ionization threshold, such a region would extend over the distance of $\sim600\,\mathrm{nm}$. Subsequently, it is necessary to trace the position of the ion with a high spatial and temporal resolution on the order of $100\,\mathrm{nm}$ and $100\,\mathrm{ns}$, respectively (compare Fig.~\ref{fig:rx_density}). An ion microscope \cite{Pfau_PRX21,Stecker_2017} is capable of directly imaging the ion’s position with a sufficient resolution as it does not rely on optical detection, thus surpassing the resolution limit of visible light. However, to avoid constant acceleration of the ion, it is necessary to compensate for electric stray fields as well as possible. Typically, related experiments reach a residual stray field level of $E_\mathrm{stray} = 0.1\,\mathrm{V\,m}^{-1}$ \cite{Engel2018}. Such a field would cause an acceleration of $a = e E_\mathrm{stray} / M$, yielding an additional velocity of $v = a t = 0.1\,\mathrm{m\,s}^{-1}$ during the calculated time span of $t = 1\,\mathrm{\mu s}$ that is below the initial velocities assumed here. Thus, nonetheless, a slowing of the ion should be observable experimentally. A more sophisticated approach would need the derivation of a master equation with a constant acceleration term due to the stray field, which is beyond the scope of this work, but an obvious extension for future work.\\
\\
\textit{Inelastic processes -} In our analysis, we have studied the cooling dynamics of the ion, which arises from elastic collisions with the atoms of the gas. However, in the case of homonuclear systems such as $^{87}\mathrm{Rb}^+$/$^{87}\mathrm{Rb}$, resonant charge exchange (RCx) can be relevant~\cite{Dalgarno_2000,Rangwala_2012}. This phenomenon consists of the charge of the ion being transferred to a neutral atom after a collision. To estimate its impact, let us first recall that two-body collisions can be divided into two groups: glancing collisions, where the particles trajectories are slightly deflected, and Langevin collisions, which can be classically represented as the two particles getting close in a spiraling motion and being scattered isotropically. In the same classical picture, Langevin collisions occur when the impact parameter of the collision is smaller than a critical value $b_c=(2C_4/E_\mathrm{col})^{1/4}$ \cite{Haerter2014}, where $C_4$ is the prefactor of the polarization potential and $E_\mathrm{col}$ is the energy of the collision. It has recently been observed in Ref.~\cite{Mahdian_2021} that RCx associated with glancing collisions can be the dominant process for collisional energies higher than $100\,\mathrm{K}\cdot k_\mathrm{B}$, leading to fast cooling of the ion (so-called swap cooling). On the other hand, for lower energies, RCx can only occur by Langevin collisions, with a cross section given by $\sigma_\mathrm{RCx}=\sigma_\mathrm{Lgv}/2$. For the $^{87}\mathrm{Rb}^+$/$^{87}\mathrm{Rb}$ system with $E_\mathrm{col}=1\,\mathrm{mK}\cdot k_\mathrm{B}$, $\sigma_\mathrm{RCx}$ is about 10 times smaller than the elastic cross section \cite{Dalgarno_2000,Haerter2014} and remains significantly lower than the latter even for $E_\mathrm{col}=2\,\mathrm{\mu K}\cdot k_\mathrm{B}$, where it reaches $1/3$ of the elastic cross section. Hence, resonant charge exchange is never dominant in the energy range of the ion. However, note that the previous reasoning is based on a semi-classical analysis, which is accurate down to $1\,\mathrm{mK}\cdot k_\mathrm{B}$. A more accurate estimation for lower energies requires studies at the quantum level that are not yet available.\\
Similar arguments can be applied to three-body recombination processes that lead to the formation of molecules. In this regard, experiments involving a trapped $^{87}\mathrm{Rb}^+$ ion immersed in an ultracold cloud of $^{87}\mathrm{Rb}$ atoms \cite{Denschlag_PRL12} showed that the three-body recombination rate is on the order of a second for values of the atomic density, comparable to the ones we considered in this work. Although our study does not involve a trap and our collision energies are lower than the ones considered in the aforementioned experiment, we can assume the formation of molecules not to play a significant role due to the short time scales in which we expect the cooling and pinning dynamics to take place.

\section{Summary and conclusions}
\label{sec:summary}
We studied the behavior of an ion moving in an ultracold bosonic gas with an initial momentum resulting from an ionization process.
To this end, in Sec.~\ref{sec:ME}, we derived the quantum master equation reported in Eq.~\eqref{eq:complete_ME}. Based on this equation, we computed the differential equations for the expectation value of the squared momentum [see Eq.~\eqref{eq:p2_equation}] and the expectation value of the position and momentum [see Eq.~\eqref{eq:p_and_x})]. We numerically solved these differential equations for different values of initial momentum $k_0$, condensate density $n_0$, and atom-ion scattering length $a_\mathrm{ai}$ and showed the corresponding results in Sec.~\ref{sec:results}. As a key observation, we demonstrated that the ion temperature defined as $T_\mathrm{ion}=\langle\hat{p}^2\rangle/(2Mk_\mathrm{B})$ decays in time. We quantified this behavior by defining the FDHM (i.e., full duration at half maximum) as the time required for $T_\mathrm{ion}$ to halve. Interestingly, we found a linear dependence of the FDHM on the mean distance between the bosons. Expanding on our key point of short cooling times, we found that the FDHM is almost independent of the initial temperature of the ion (Fig.~\ref{fig:WHM}a), whereas it is noticeably affected by the density of the condensate (Fig.~\ref{fig:WHM}b). Similarly, we observed that the ion's velocity drops by nine orders of magnitude in a time that is independent of the ion's initial velocity (Fig.~\ref{fig:first_moments}), which we attribute to incoherent dynamics of the phonon modes as a consequence of the enhancement of the friction coefficient (Fig.~\ref{fig:friction}). In conclusion, our study predicts the cooling and pinning of the ion due to its interaction with the surrounding ultracold bosonic gas. Moreover, we observed a substantial robustness of the results against the parameters involved. These findings are relevant in view of the upcoming experiments discussed in Sec.~\ref{sec:exp}, as the time and length scales of the ion's dynamics are compatible with the expected experimental resolution.

\section*{Acknowledgements}
This work is supported by the project NE 1711/3-1 of the Deutsche Forschungsgemeinschaft (DFG). Moreover, support by the Cluster of Excellence ``CUI: Advanced Imaging of Matter'' of the DFG — EXC 2056 — project ID 390715994 is acknowledged.

\appendix

\section{\addtwo{Validity of the Lamb-Dicke approximation}}
\label{app:LD}
\begin{figure}
    \centering
    \includegraphics[width=.45\textwidth]{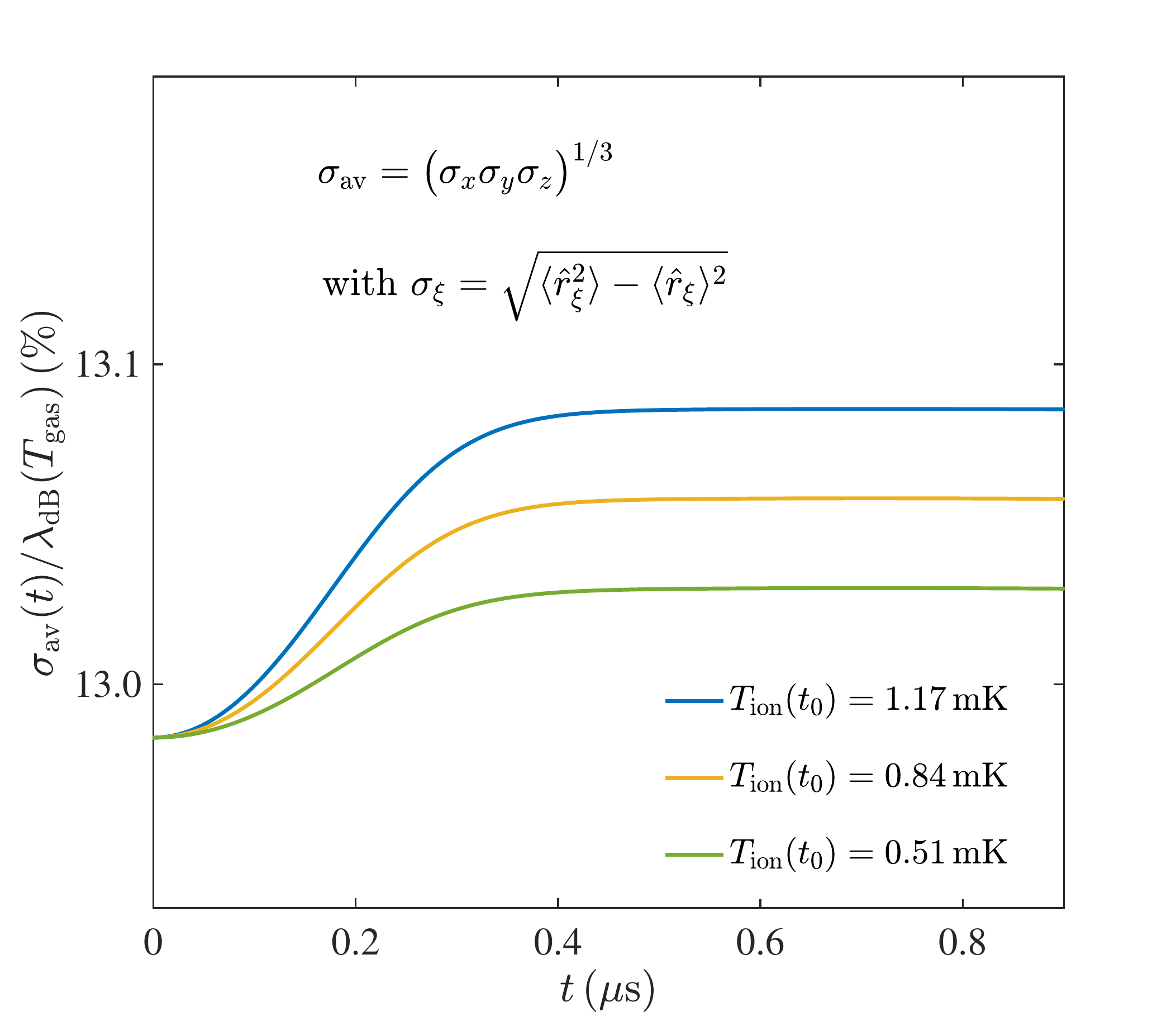}
    \caption{\addtwo{Average spatial width of the ion as a function of time as a percentage of the atomic de Broglie wavelength $\lambda_\mathrm{dB}(T_\mathrm{gas})$ with $T_\mathrm{gas}=1\,\mathrm{nK}$ and $n_0=2\cdot10^{14}\,\mathrm{cm}^{-3}$. The three lines correspond to the three initial temperatures considered in Fig~\ref{fig:Ttot}(a).}}
    \label{fig:LD_check}
\end{figure}

\addtwo{The master equation derived in Sec.~\ref{sec:ME} relies on the Lamb-Dicke approximation. We recall that the latter results in the expansion in Eq.~\eqref{eq:LD_expansion} and it is based on the assumption that the average wavelength of the atoms in the bath is much larger than the width of the ion. Here, we discuss the fulfillment of such a condition during the evolution of the system in question. Of course, the assumption has to hold regardless of the choice of the initial state. Since the spatial extension of the ion wave function at the initial time must fulfill the approximation as well, this imposes a condition on the temperature of the gas. \\
Let us consider, for example, the ionization process via a Rydberg state. In order to hold, the Lamb-Dicke approximation imposes that $\lambda_\mathrm{dB}(T_\mathrm{gas})\gg\sigma_\mathrm{av}$, where $\sigma_\mathrm{av}$ is the geometric average of $\sigma_\xi=\sqrt{\langle\hat{r}_\xi^2\rangle - \langle\hat{r}_\xi\rangle^2}$ along the three directions and represents the width of the ion wave packet (i.e., the standard deviation), while $\lambda_\mathrm{dB}(T_\mathrm{gas})$ is the de Broglie wavelength of the bosons in the gas at temperature $T_\mathrm{gas}$. A rough estimate is given by considering the ion to be in the ground state of the harmonic trap in Eq.~\eqref{eq:rho_xi}. 
Thus, we have at the initial time $\sigma_\mathrm{av}=\sqrt{\hbar/(2M\omega_\mathrm{av})}$ with $\omega_\mathrm{av} = (\omega_\mathrm{x} \omega_\mathrm{y} \omega_\mathrm{z})^{1/3}$ being the geometrical average of the harmonic trap frequencies. For a homonuclear system and with the trap frequencies shown in Tab.~\ref{tab:experimental_values}, we get the following condition on the gas temperature
\begin{equation}
    T_\mathrm{gas}\ll\frac{4\pi\hbar}{k_\mathrm{B}}\frac{M}{m}\omega_\mathrm{av},
    \label{eq:Tgas_condition}
\end{equation}
which yields a value of $T_\mathrm{gas}$ on the order of nK. At such a low temperature, the exponential weights in the sum of Eq.~\eqref{eq:x02_formula} barely affect the value of $\langle\hat{r}_\xi^2(t_0)\rangle$. For this reason, we can safely use Eq.~\eqref{eq:Tgas_condition} as a condition for our system to be in the Lamb-Dicke regime at the initial time. Note that the latter statement could be violated for ion-atom systems with a different mass ratio.\\
In the case of ionization via an ultrashort laser pulse, the condition for the validity of the Lamb-Dicke approximation can be simply verified by comparing the de Broglie wavelength of the gas with the value of the laser beam waist $w_0$. Considering $w_0=1\,\mu\mathrm{m}$, as it is expected in future experiments, we can impose a condition on $T_\mathrm{gas}$. Similarly, the required value is on the order of nK.\\
The validity of the Lamb-Dicke approximation at later times, however, has to be monitored numerically by solving the equations for the first [Eq.~\eqref{eq:p_and_x}] and second order moments [Eq.~\eqref{eq:p2_equation} and \eqref{eq:r2_c_equations}]. In Fig.~\ref{fig:LD_check}, we show the time evolution of the spatial width of the ion $\sigma_\mathrm{av}$ for three different initial temperatures. As it can be observed, the ratio between $\sigma_\mathrm{av}$ and the de Broglie wavelength is always on the order of one tenth, confirming that the Lamb-Dicke approximation is rather well justified.}

\section{Details on the master equation}
\label{app:ME}
For the sake of completeness, let us retrace the main steps of the derivation starting from the von Neumann equation for the density matrix of the total system (ion plus bath) $\hat{\chi}(t)$:
\begin{equation}
    \frac{d}{dt}\hat{\chi}(t)=-\frac{i}{\hbar}\comm{\hat{H}}{\hat{\chi}(t)}.
    \label{eq:von_Neumann}
\end{equation}
Following the standard quantum-optical approach (see, e.g., Ref.~\cite{CarmichaelQOBook}), we write the density matrix in the interaction picture as
\begin{equation}
    \Tilde{\chi}(t)=\hat{U}^\dag(t)\hat{\chi}(t)\hat{U}(t)
    \label{eq:chi_tilde}
\end{equation}
with
\begin{equation}
    \hat{U}(t)=\mathrm{exp}\bigg[-\frac{i}{\hbar}\Big(\hat{H}_\mathrm{ion}+\hat{H}_\mathrm{bath}\Big)t\bigg].
    \label{eq:U}
\end{equation}
Recalling the definition of the total Hamiltonian $\hat{H}=\hat{H}_\mathrm{ion}+\hat{H}_\mathrm{bath}+\hat{H}_\mathrm{int}$, we obtain the following equation
\begin{equation}
    \frac{d}{dt}\Tilde{\chi}(t)=-\frac{i}{\hbar}\comm{\Tilde{H}_\mathrm{int}}{\Tilde{\chi}(t)},
    \label{eq:von_Neumann_tilde}
\end{equation}
whose formal solution reads
\begin{equation}
    \Tilde{\chi}(t)=\Tilde{\chi}(0)-\frac{i}{\hbar}\int_0^tdt'\,\comm{\Tilde{H}_\mathrm{int}(t')}{\Tilde{\chi}(t')}.
    \label{eq:formal_solution}
\end{equation}
Here, $\Tilde{H}_\mathrm{int}$ is the interaction Hamiltonian in the interaction picture.\\
We now insert Eq.~\eqref{eq:formal_solution} in Eq.~\eqref{eq:von_Neumann_tilde} and we get
\begin{equation}
    \begin{split}
        \frac{d}{dt}\Tilde{\chi}(t)=&-\frac{i}{\hbar}\comm{\Tilde{H}_\mathrm{int}(t)}{\Tilde{\chi}(0)}\\
        &-\frac{i}{\hbar^2}\int_0^tdt'\,\comm{\Tilde{H}_\mathrm{int}(t)}{\comm{\Tilde{H}_\mathrm{int}(t')}{\Tilde{\chi}(t')}}.
    \end{split}
    \label{eq:chi_tilde_equation}
\end{equation}
In order to proceed, we assume that at the initial time $t=0$ the system and the bath are not correlated. This allows the total density matrix to be factorized as $\Tilde{\chi}(0)=\Tilde{\rho}(0)\otimes\hat{B}_0$, where $\Tilde{\rho}(0)=\hat{\rho}(0)$ is the initial density matrix of the ion, while $\hat{B}_0$ is the initial density matrix of the Bose gas at thermal equilibrium
\begin{equation}
\label{eq:B0}
    \hat{B}_0=\frac{e^{-\beta\big(\hat{H}_\mathrm{bath}-\mu_\mathrm{G}\hat{N}\big)}}{\mathcal{Z}},\quad \mathcal{Z}=\mathrm{Tr}_\mathrm{b}\bigg\{e^{-\beta\big(\hat{H}_\mathrm{bath}-\mu_\mathrm{G}\hat{N}\big)}\bigg\},
\end{equation}
where $\beta=1/(k_\mathrm{B}T_\mathrm{gas})$, $\hat{N}$ is the bath number operator and the chemical potential of the gas $\mu_\mathrm{G}$ is zero for a Bose gas below the critical temperature of condensation.\\
Note that the same assumption was made in Ref.~\cite{Oghittu_PRA21}, where a Paul-trapped ion immersed in an ultracold gas was considered. In that case, the assumption was well justified, as the ion and the gas are typically prepared separately in experiments, and no interaction occurs before they are brought together. In the present case, we note that part of the simulations will refer to a scenario where the ion is created after ionizing one of the atoms in the gas. However, the interaction between the gas atoms is weaker and of short-range nature compared to the atom-ion polarization potential. Given the fact that the ionization process occurs on a time scale much shorter than every other time scale in our theoretical treatment, we can reasonably assume that at the very initial moment of the ion generation the interaction between the ion and the bath is weak. Only subsequently, it becomes stronger, but in such a way that the gas state is not significantly altered.\\
Now, we trace out the bath degrees of freedom from Eq.~\eqref{eq:chi_tilde_equation} obtaining an equation for the reduced density matrix of the ion
\begin{equation}
    \frac{d}{dt}\Tilde{\rho}(t)=-\frac{1}{\hbar^2}\int_0^tdt'\,\mathrm{Tr}_\mathrm{b}\Big\{\comm{\Tilde{H}_\mathrm{int}(t)}{\comm{\Tilde{H}_\mathrm{int}(t')}{\Tilde{\chi}(t')}}\Big\},
    \label{eq:rho_tilde_eq}
\end{equation}
and we finally perform the Born and Markov approximations. The Born approximation relies on the fact that the coupling between the ion and the bath is weak and that the bath is very large. Therefore, the factorization $\Tilde{\chi}(t')\simeq\Tilde{\rho}(t')\otimes\hat{B}_0$ is assumed to be valid at all times $t'$. Instead, the Markov approximation is based on the assumption that the dynamics of the bath is much faster than the dynamics of the ion. It consists of the replacement $\Tilde{\rho}(t')\to\Tilde{\rho}(t)$.\\
We then get to the so-called Redfield equation for the reduced ion density matrix in the interaction picture:
\begin{equation}
   \frac{d}{dt}\Tilde{\rho}(t)=-\frac{1}{\hbar^2}\int_0^tdt'\,\mathrm{Tr}_\mathrm{b}\Big\{\comm{\Tilde{H}_\mathrm{int}(t)}{\comm{\Tilde{H}_\mathrm{int}(t')}{\Tilde{\rho}(t)\otimes\hat{B}_0}}\Big\}.
\end{equation}
From this equation, we can explicitly substitute $\Tilde{H}_\mathrm{int}$ and transform back to the Schr\"{o}dinger picture. After tracing out the bath degrees of freedom, we get to the impurity master equation provided in Eq.~\eqref{eq:BEC_ME}, where we performed the change of variable $\tau=t-t'$ and $\hat{\mathbf{r}}(t,\tau)=\hat{U}(t)\hat{U}^\dag(t-\tau)\hat{\mathbf{r}}\hat{U}(t-\tau)\hat{U}^\dag(t)$.\\
Here, we report for the interested reader the complete master equation in the Born-Markov and Lamb-Dicke approximation:
\begin{equation}
    \begin{split}
        \frac{d}{dt}\hat{\rho}(t)=-\frac{i}{\hbar}&\comm{\hat{H}_\mathrm{S}}{\hat{\rho}(t)}\\
        -\sum_{\xi=x,y,z}\Big\{&\lambda_\xi(t)\comm{\hat{r}_\xi}{\hat{\rho}(t)\hat{p}_\xi}-\lambda^*_\xi(t)\comm{\hat{r}_\xi}{\hat{p}_\xi\hat{\rho}(t)}\\
        +&\phi_\xi(t)\Big(\comm{\hat{r}_\xi}{\hat{r}_\xi\hat{\rho}(t)}-\comm{\hat{r}_\xi}{\hat{\rho}(t)\hat{r}_\xi}\Big)\Big\},
    \end{split}
\label{eq:complete_ME}
\end{equation}
where we have introduced the following functions:
\begin{equation}
    \begin{split}
        \lambda_\xi(t)=\sum_{\mathbf{q}}\Omega_q^2\,q_\xi^2\bigg\{&\frac{2n_q}{m\omega_q^2}\big[\cos(\omega_qt)+\omega_qt\sin(\omega_qt)-1\big]\\
        &+\frac{1}{m\omega_q^2}\Big[e^{i\omega_qt}\big(1-i\omega_qt\big)-1\Big]\bigg\},\\
        \phi_\xi(t)=\sum_{\mathbf{q}}\Omega_q^2\,q_\xi^2\bigg[&(2n_q+1)\frac{\sin(\omega_qt)}{\omega_q}\bigg].
    \end{split}
\end{equation}
The equations in Eq.~\eqref{eq:p2_equation}, \eqref{eq:r2_c_equations} and \eqref{eq:p_and_x} are computed by explicitly calculating the expectation value of the corresponding observables with Eq.~\eqref{eq:complete_ME} and by making use of the canonical commutation relations between the position and momentum operators.

\section{Alternative squared momentum derivation}
\label{sec:alternative_derivation}
The same equation for the squared momentum in Eq.~\eqref{eq:p2_equation} can be derived with a different approach. In particular, we can calculate the variation in the ion's energy due to the presence of the gas. Specifically, to second order in the perturbative expansion we have 
\begin{equation}
    \begin{split}
        \frac{d\langle\hat{H}_\mathrm{ion}(t)\rangle}{dt}=&\frac{i}{\hbar}\Big\langle\comm{\Tilde{H}_\mathrm{int}(t)}{\hat{H}_\mathrm{ion}}\Big\rangle\\
        -\frac{1}{\hbar^2}&\int_0^t dt'\,\Big\langle\comm{\Tilde{H}_\mathrm{int}(t')}{\comm{\Tilde{H}_\mathrm{int}(t')}{\hat{H}_\mathrm{ion}}}\Big\rangle,
    \end{split}
    \label{eq:cooling_rate}
\end{equation}
where the average value $\langle\dots\rangle$ has to be intended as the trace over a density matrix. By choosing the total system density matrix as the one we defined for the derivation of the master equation, it is straightforward to show that the first term on the right hand side of Eq.~\eqref{eq:cooling_rate} vanishes due to the odd number of bath operators, while the second gives rise to terms proportional to $n_\mathbf{q}$ and $(n_\mathbf{q}+1)$. After transforming back to the Schr\"{o}dinger picture, performing the time integrals and the Lamb-Dicke approximation, one can retrieve Eq.~\eqref{eq:p2_equation}.

\section{\addone{Dynamics for low initial ion temperatures}}
\label{app:low_T}
\begin{figure}
    \centering
    \includegraphics[width=.45\textwidth]{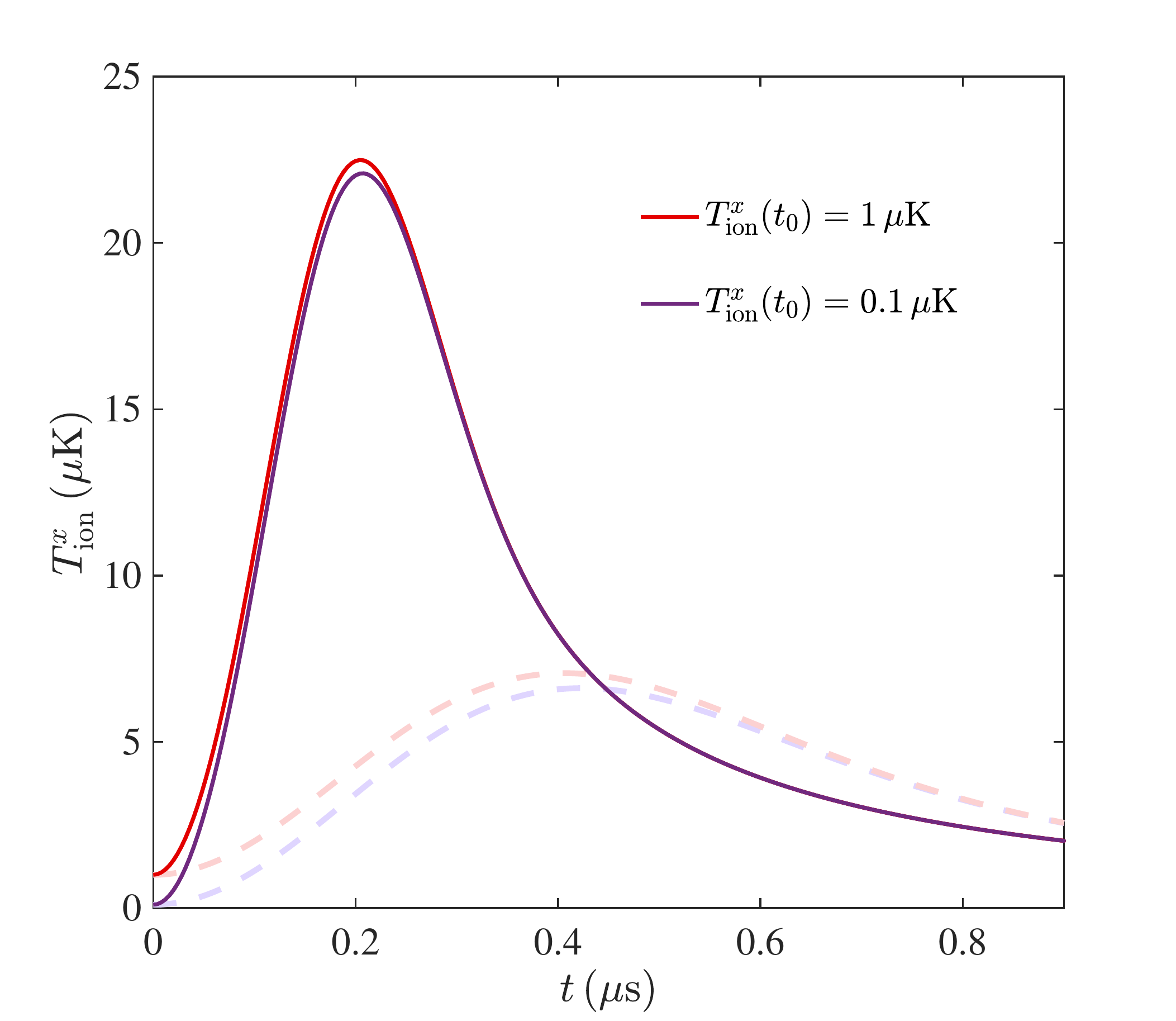}
    \caption{Ion temperature $T_\mathrm{ion}^x=\langle\hat{p}_x^2\rangle/(2Mk_\mathrm{B})$ as a function of time for $n_0=2\cdot10^{14}\,\mathrm{cm}^{-3}$ (solid lines) and $n_0=2\cdot10^{13}\,\mathrm{cm}^{-3}$ (light dashed lines). The initial ion temperatures correspond to $T_\mathrm{ion}^x=1\,\mathrm{\mu K}$ (red) and $T_\mathrm{ion}^x=0.1\,\mathrm{\mu K}$ (purple).}
    \label{fig:low_T}
\end{figure}

\addone{Here we discuss the time evolution of the ion temperature obtained for initial values in a regime comparable to $E^\star/k_\mathrm{B}=79\,\mathrm{nK}$ and one order of magnitude higher. The results are shown in Fig.~\ref{fig:low_T}. Interestingly, the ion is heated up at short times, meaning that the expectation value of its kinetic energy increases. At later times, the ion temperature exhibits a maximum. This is positioned around $\sim0.2\,\mu\mathrm{s}$ for $n_0=2\cdot10^{-14}\,\mathrm{cm}^{-3}$ and $\sim0.4\,\mu\mathrm{s}$ for $n_0=2\cdot10^{-13}\,\mathrm{cm}^{-3}$. Similarly to what we observed for initial ion energies in the mK-regime, the results in Fig.~\ref{fig:low_T} only slightly depend on the initial temperature, while they are noticeably affected by the density of the condensate. In particular, a lower density (light dashed lines) corresponds to a shift towards larger times and flattening of the maximum. The behavior of $T_\mathrm{ion}$ at short times can be attributed to the long-range and attractive character of the atom-ion interaction generated after the ionization process. Contrarily to a neutral impurity, which interacts with a particle of the bath only when this is at its same position, the increased range of the polarization potential causes the ion to heat up due to the surrounding polarized bath atoms within a radius given by $R^\star$. This dynamical behavior continues until the frequency of collisions with the atoms in the bath is large enough to cool down the moving ion. We note that a similar initial heating, although with a lower peak, is also observed with initial temperatures in the mK-regime. However, the scale of temperatures in the main plot of Fig.~\ref{fig:Ttot} does not allow this peaks to be appreciated. It is also important to remark that no maximum is observed in the expectation value of the ion momentum $\langle\hat{p}_x(t)\rangle$, meaning that the heating of the ion at short time does not correspond to an acceleration.}


\bibliography{references}

\end{document}